\documentclass[journal]{IEEEtran}
\usepackage{amsmath}
\usepackage[ruled]{algorithm2e} 
\usepackage{caption}
\usepackage{rotating}
\usepackage{subcaption}
\usepackage{pgfplots,pgfplotstable}
\usetikzlibrary{pgfplots.groupplots}
\usepackage{tikz}
\usepackage{lipsum,adjustbox}
\usetikzlibrary{calc,positioning}
\pgfplotsset{compat=newest}

\newcommand{\eg}[0]{\textit{e.g.},~}
\newcommand{\ie}[0]{\textit{i.e.},~}

\begin{document}
	%
	\title{Deep Tone Mapping Operator for High Dynamic Range Images}
	
	\author{Aakanksha Rana*,
		Praveer Singh*, Giuseppe Valenzise, Frederic Dufaux, Nikos Komodakis, Aljosa Smolic
		\thanks {A. Rana and A. Smolic are with V-SENSE, Trinity College Dublin, Ireland}
		\thanks{G. Valenzise  and  F. Dufaux are with Laboratoire des Signaux et Syst\'emes, CNRS, CentraleSupelec, Universit\'e Paris-Sud.}
		\thanks{P. Singh and N. Komodakis are with LIGM/IMAGINE,
			Ecole des Ponts ParisTech, Universit\'e Paris-Est.}
		\thanks{*Equal contributors. This publication has emanated from research conducted with the financial support of Science Foundation Ireland (SFI) under the Grant Number 15/RP/2776. The work presented in this document was also supported by BPIFrance and R\'egion Ile de France, in the framework of the FUI 18 Plein Phare project. We gratefully acknowledge the support of NVIDIA Corporation with the donated GPU used for this research.}
		
	}
	
	\maketitle
	
\begin{abstract}
A computationally fast tone mapping operator (TMO) that can quickly adapt to a wide spectrum of high dynamic range (HDR) content is quintessential for visualization on varied low dynamic range (LDR) output devices such as movie screens or standard displays. Existing TMOs can successfully tone-map only a limited number of HDR content and require an extensive parameter tuning to yield the best subjective-quality tone-mapped output. In this paper, we address this problem by proposing a fast, parameter-free and scene-adaptable deep tone mapping operator (DeepTMO) that yields a high-resolution and high-subjective quality tone mapped output. Based on conditional generative adversarial network (cGAN), DeepTMO not only learns to adapt to vast scenic-content (\eg outdoor, indoor, human, structures, etc.) but also tackles the HDR related scene-specific challenges such as contrast and brightness, while preserving the fine-grained details. We explore 4 possible combinations of Generator-Discriminator architectural designs to specifically address some prominent issues in HDR related deep-learning frameworks like blurring, tiling patterns and saturation artifacts. By exploring different influences of scales, loss-functions and normalization layers under a cGAN setting, we conclude with adopting a multi-scale model for our task. To further leverage on the large-scale availability of unlabeled HDR data, we train our network by generating \textit{targets} using an objective HDR quality metric, namely Tone Mapping Image Quality Index (TMQI). We demonstrate results both quantitatively and qualitatively, and showcase that our DeepTMO generates high-resolution, high-quality output images over a large spectrum of real-world scenes. Finally, we evaluate the perceived quality of our results by conducting a pair-wise subjective study which confirms the versatility of our method.
\end{abstract}
	

	\section{Introduction} \label{sec:intro}
Tone mapping is a prerequisite in the high dynamic range (HDR) imaging~\cite{Pardo,Tmqi2,Gommelet, rana-icme17} pipeline to print or render HDR content for low dynamic range displays. With the unprecedented demands of capturing/reproducing scenes in high-resolution and superior quality, HDR technology is growing rapidly~\cite{dufaux2016book,bookHDR, rana-icip17}.
Although HDR display systems have advanced in the last few decades (for \eg Sim2, Dolby Vision, etc), they still necessitate some sort of tone mapping operation because of limited technical capabilities of the materials used in these displays. Additionally, due to high manufacturing costs, the absolute majority of screens still have limited dynamic range and rely largely on Tone Mapping Operators (TMOs) for desired top-quality presentation.

Several TMOs have been designed over the last two decades, promising the most faithful representation of real-world luminosity and color gamut for high-quality output. However, in practice, such TMOs are limited to successfully tone map only limited number of HDR images due to their parametric sensitivity~\cite{ledda2005evaluation,Gabriel}. For instance, a TMO capable of mapping a bright daytime scene might not map a dark or evening scene equally well. In fact, one needs to manually tweak in an extensive parametric space for every new scene, in order to achieve the best possible results while using any such TMO. Thus, the entire process of finding the most desirable high-resolution tone-mapped output is not only slow, tedious and expensive, but is almost impractical when there is a large variety of HDR content being generated from numerous capturing devices.

This raises a natural question whether a more \textit{adaptive} tone mapping function can be formulated which can quickly alter itself to wide variability in real-world HDR scenes to reproduce the best subjective quality output without any perceptual damage to its content on a high-resolution display. With the recent success of deep learning \cite{krizhevsky2012imagenet} and wide scale availability of HDR data, it is now possible to learn a model with such complex functionalities for effective tone mapping operation. 

In this paper, we propose an end-to-end deep learning (DL) based tone-mapping operator (DeepTMO) for converting any given HDR scene into a tone-mapped LDR output which is of high resolution [1024x2048] and superior subjective quality. Based upon a conditional generative adversarial network (cGAN)~\cite{gan,cgan}, the DeepTMO model directly inputs 32-bit~\textit{linear} HDR content and reproduces a realistically looking tone-mapped image, aiming to mimic the original HDR content under a limited range [0-255]. DeepTMO is trained to cater a wide range of scenic-content for \eg indoor/outdoor scenes, scenes with structures, human faces, landscapes, dark and noisy scenes, etc.


The motivation for generative adversarial networks (GAN) in the DeepTMO design stems from their tremendous success in several image-to-image translation studies~\cite{pix2pix}. Such models have shown to overcome the problem of spatially blurred-out resulting images with a simple $L_1/L_2$ loss function. Furthermore, instead of optimizing parameters for a given TMO~\cite{ranammsp} for a particular scene~\cite{kurt,Tmqi2}, our objective is to design a model which is \textit{adaptable} to different scenes-types (such as day/night, outdoor/indoor, etc.), thus encompassing all their desired characteristics. Altogether, this is difficult for a naive loss-function to satisfy. Moreover, designing such a cost function is quite complex~\cite{rana2014}, and needs expert knowledge. Therefore, we overcome this challenge by \textit{learning} an `adversarial loss' that encapsulates all the desired features from all ideal tone-mapped images by using the underlying training data; thereby eradicating the need of manually handcrafting such a loss function. 

GANs are capable to generate better quality images compared to the state-of-the-art models, however, there are still some prominent issues such as tilling patterns, local blurring and saturated artifacts (see Fig.~\ref{fig:MSDMSGimpact} (a)). To handle these problems in a high-resolution output image, we explore the DeepTMO architectural design by comparing the single-scale and multi-scale variants of both generator and discriminator. We subsequently showcase how a multi-scale version of the generator-discriminator architecture helps in predicting artifact-free tone mapped images, which are both structurally consistent with input HDR and simultaneously preserve fine-grained information recovered from different scales. 

The DeepTMO model is effectively a multi-scale architecture having a 2-scale generator and a 2-scale discriminator, both of which are conditioned on the \textit{linear} HDR input. Both generator and discriminator compete with each other. The generator is trying to fool discriminator by producing high subjective quality tone mapped images for the given input HDR, while the discriminator trying to discriminate between real and synthetically generated HDR-LDR image pairs. Our basic discriminator architecture is similar to PatchGAN~\cite{li2016precomputed,ledig2016photo} which classifies patches over the entire image and averages over all of them to yield the final image score. Similarly our basic generator architecture comprises of an encoder-decoder network where the input HDR is given first to an encoder resulting in a compressed representation which is then passed to the decoder yielding finally a tone mapped image. 

To train our model, we accumulate our dataset from freely available HDR image sources. Ideally, the training dataset should be created through a subjective evaluation considering all possible tone mapping operators for all available HDR scenes. However, conducting such a subjective evaluation is highly cumbersome and unfeasible. Thus, it necessitates the requirement of an objective quality assessment metric which can quantify the tone mapping performance of each TMO for any given scene. For our task, we select a well known metric namely Tone Mapped Image Quality Index (TMQI). We first rank 13 widely used TMOs using the TMQI metric for each HDR input. We then select the topmost scoring tone-mapped image as our \textit{target} output.

In a nutshell, we 
\begin{enumerate}
	\item propose a fast, parameter-free DeepTMO, which can generate high-resolution and foremost subjective quality tone-mapped outputs for a large variety of \textit{linear} HDR scenes, including indoor, outdoor, person, structures, day and night/noisy scenes.
	\item explore 4 possible cGANs network settings: (a) Single-scale-Generator (Single-G) and Single-scale-Discriminator (Single-D), (b) Multi-scale-Generator (Multi-G) and Single-D, (c) Single-G and Multi-scale-Discriminator (Multi-D), (d) Multi-G and Multi-D, thus discussing the influence of scales and finally proposing a multi-scale generator-discriminator model for our problem. 
	\item detail the impact of different loss functions and normalization layers while elaborating how each step helps in improving the overall results by tackling different artifacts. 
	\item provide quantitative and qualitative comparison of our model with best tone mapped outputs over 105 images and also validate our technique through a pair-wise subjective study. 
\end{enumerate}
	\section{Related Work}\label{sec:rw}

HDR imaging technology has been a subject of interest over the past decades, inspiring to capture and reproduce a wide range of colors and luminous intensities of the real world on a digital canvas. Normally, the information stored in HDR content is represented using a 32-bit floating point format. But to cope with conventional displays, such scenes are often tone-mapped to an LDR format with available TMOs. A great variety of TMOs addressing different perceptual objectives have been proposed in the past years. In the following, we give a quick review of the tone mapping literature and then would touch upon various deep learning techniques for HDR imaging.

\subsection{Tone Mapping Operators for HDR Content}
TMOs have been widely explored in the literature, principally based upon how they handle the contrast, color and luminosity in a given HDR image~\cite{Banterle}. However, they have been classified into several categories under different sets of criteria~\cite{bookHDR,dufaux2016book}. Primarily, they have been grouped into \textit{global} and \textit{local} approaches, relying on how these mapping functions operate on an image. The global methods such as~\cite{ward97,drago03,schlick94} apply the same compression function to all the pixels of an image. For the \textit{local} techniques such as~\cite{tumblin99,patt02,chiu96}, a tone-mapped pixel depends on the values of its neighboring pixels. Even though global approaches are faster to compute, their resulting LDR outputs do not maintain adequate contrast in the images; thus the scene appears somewhat washed out. The local tone mapping functions, conversely, do not face these issues and are generally capable enough of handling contrast ratios, meanwhile preserving local details. However, these operators result in some prominent `halo' effects around the high frequency edges, thereby giving unnatural artifacts in the scenes. Another category of TMOs~\cite{durand02,mantiuk06,fattal02} includes designs which are inspired from the human visual system, can models the attributes such as adaptation with time, and can discriminate at high contrast stimuli and gradient sensitivities. 
Nonetheless, all these existing TMOs have been designed to target independently, multiple different objectives~\cite{bookHDR,Gabriel}, such as simulating human visual properties, honest reproduction of scenes, best subjective preference or even for computer vision applications~\cite{RanaTmm}. However, in our work, we mainly focus towards designing a TMO aiming for ``best subjective quality output''.  

Several small scale perceptual studies have been performed using varied criteria such as with reference or without reference~\cite{cadik08,ledda2005evaluation,tmqi} to compare these classical and newly developed TMOs for different perceptual objectives. Even though these subjective studies are ideal to analyze TMO's performance, the process is bounded to use a limited number of content and TMOs due to practical considerations. As an alternate solution, objective metrics such as ~\cite{tmqi,fsitm} have been proposed to automate the evaluation. TMQI is a state-of-the-art objective metric and has been widely used for several TMO optimization studies~\cite{Tmqi2,kurt}. It assesses the quality of images on 1) structural fidelity which is a multi scale analysis of the signals, and 2) naturalness, which is derived using the natural image statistics. Both these crucial properties of human perception are combined to define a subjective quality score. 

\paragraph{Learning-based methods}
Parametric sensitivity of hand-crafted TMOs is a well-known phenomenon which impacts the subjective quality of the resulting output. As a result, this emphasizes `scene-dependence' of such tone mapping designs \ie for a given subjective quality task, TMOs have to be fine tuned for each individual scene type. To this end, some optimization based tone mapping frameworks~\cite{Tmqi2,kurt} have been designed where the parameters of a specific TMO are optimized for a given image. However, the parameter fine-tuning process for each scene separately is time consuming and limits its real-time applicability. Additionally, it somehow questions the `automatic' nature of tone mapping~\cite{Gabriel} for their applicability on a wide variety of real-world scenes. 

\subsection{ CNNs for HDR Scenes}

Recently, CNNs have been utilized extensively for multiple HDR imaging tasks such as reconstructing HDR using a single-exposure LDR \cite{eilertsen2017hdr,ExpandNet,AKinoshita,BMoriwaki}, predicting and merging various high and low exposure images for HDR reconstruction \cite{deepreverse} or yielding HDR outputs from dynamic LDR inputs \cite{dynamichdr}. CNNs have also been modeled to learn an input-output mapping as done for de-mosaicking and de-noising by \cite{gharbi2016deep} or learning an efficient bilateral grid for image enhancement \cite{chen2016bilateral}.~\cite{Gharbi} have recently proposed a deep bilateral tone mapper, but it works only for 16-bit linear images and not for conventional 32-bit HDR images. A recent work~\cite{deeptmo} addresses the end-to-end tone mapping problem where the model is trained for a given scene. This is somewhat similar approach to parameter-tuning where the model is calibrated for only one given scene at a time. Therefore, the problem of designing a fast, parameter-free, end-to-end TMO which can effectively tone map wide variety of real-world high-resolution content for high quality display in real time, still holds relevance.

As observed in the past CNN studies, the quality of resulting output depends heavily on the choice of the loss function. Formulating a loss function that constrains the CNN to yield sharp, top quality tone-mapped LDR from their corresponding linear-valued HDR is complex and an ill posed problem. Our work doesn't encounter such issues as we utilize a GAN based architecture.

\subsection{Generative Adversarial Networks}
GANs~\cite{gan} have attended lots of attention owing to their capability of modeling the underlying target distribution by forcing the predicted outputs to be as indistinguishable from the target images as possible. While doing this, it implicitly learns an appropriate loss function, thus eliminating the requirement of hand crafting one by an expert. This property has enabled them to be utilized for wide variety of image processing tasks such as super-resolution \cite{ledig2016photo}, photo-realistic style-transfer \cite{johnson2016perceptual} and semantic image in-painting \cite{yeh2016semantic}. 
For our task, we employ GAN under a conditional setting, commonly referred as cGAN~\cite{cgan}, where the generated output is conditioned on the input image. Recently, cGAN based frameworks have been designed for the inverse problem of generating HDR images from single LDR images~\cite{Lee_2018_ECCV,Lee_HDRi}.

One distinctive feature of cGAN frameworks is that they learn a structured loss where each output pixel is conditionally dependent on one or more neighboring pixels in the input image. Thus, this effectively constrains the network by penalizing any possible structural difference between input and output This property is quite useful for our task of tone-mapping where we only want to compress the dynamic range of an HDR image, keeping the structure of the output similar to the input HDR. For this specific reason, cGANs have been quite popular for image-to-image translation tasks, where one representation of a scene is automatically converted into another, given enough training pairs \cite{pix2pix} or without them under unsupervised settings \cite{cyclegan,liu2017unsupervised, tung2017adversarial}. However, a major limitation of using cGANs is that it is quite hard to generate high resolution images due to training instability and optimization issues. The generated images are either blurry or contain noisy artifacts such as shown in Fig.~\ref{fig:MSDMSGimpact} (a). In~\cite{crn}, motivated from perceptual loss \cite{johnson2016perceptual}, the authors derive a direct regression loss to generate high-resolution $2048 \times 1024$ images, but their method fails to preserve fine-details and textures. In~\cite{pix2pixhd}, authors have recently shown significant improvement on the quality of high-resolution generated outputs through a multi-scale generator-discriminator design. A similar work for converting HDR to LDR using GANs~\cite{patel} has also appeared recently where authors oversimplifies the tone-mapping problem by testing only on small 256x256 image crops. Essentially, such an approach may not substantially capture the full luminance range present in HDR images, and thereby overlooks the basic goal of TMO by working on full dynamic range scenes. We, however, showcase our findings using their adopted architectures from~\cite{pix2pix} on $1024\times2048$ HDR images in the supplementary material.

In summary, we motivate the DeepTMO design with these given findings, and discuss the impact of scales for both generator and discriminator, while showcasing their ability to generate high-resolution tone-mapped outputs.

	\section{Algorithm}\label{sec:algo} 
\subsection{Problem Formulation}
We propose a fast DeepTMO model with the prime objective of producing high-resolution and high-quality tone-mapped images for vast variety of real-world HDR images. Ideally, our model should automatically adapt for each scene without any external parameter tuning. To this end, we propose to imbibe different desired tone mapping characteristics depending upon scene type, content, brightness, contrast etc., to yield high perceptual quality output. In the following paragraphs, we will briefly discuss the formulation of our DeepTMO model.

\paragraph{Linear Domain Input}
For our models, we directly work on linear values. We performed the scaling to [0,1] with very high-precision (32-bit float precision), thereby, not impacting the overall output brightness. This way, we could simply automate the entire pipeline by making the network learn itself from the unaltered high-dynamic information of the scene. Additionally, we also experimented with log-scaling the input HDR before performing the tone mapping operation, specifically to test for halo-effects in high-exposure regions such as the sun in Fig.\ref{fig:haloeffect}. Note that we did some experimental studies with different input normalization techniques. More details can be found in supplementary material.

\paragraph{Color Reproduction}
Classical TMOs firstly perform the dynamic range compression in the luminance channel only and then, the colors are reproduced in the post-processing stage. This partially accounts to ease the computational complexity of the tone-mapping operation. We follow a similar paradigm, employing the common methodology for color reproduction~\cite{schlick94} given as $C_{out} = \frac{C_{in}}{L_{in}} \cdot L_{out} $, where $C_{out}$ and $L_{out}$ are output color and luminance images while $C_{in}$ is the input HDR color image. 

\begin{figure}[h]
	\begin{subfigure}[t]{0.235\textwidth}
		\caption{$\mathcal{L}_1$-loss}
		\includegraphics[width=\linewidth]{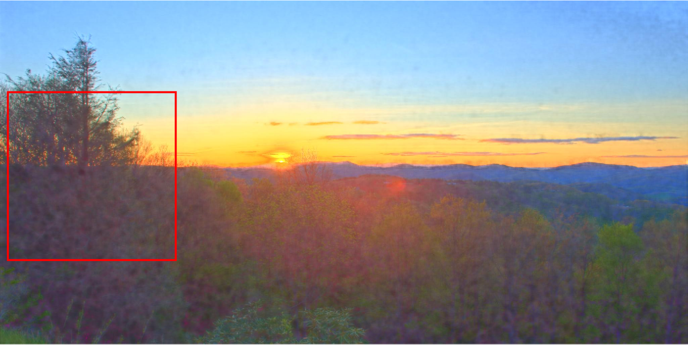}
		\label{fig:l1loss}
	\end{subfigure}
	\begin{subfigure}[t]{0.235\textwidth}
		\caption{Deep-TMO (Single-Scale)}
		\includegraphics[width=\linewidth]{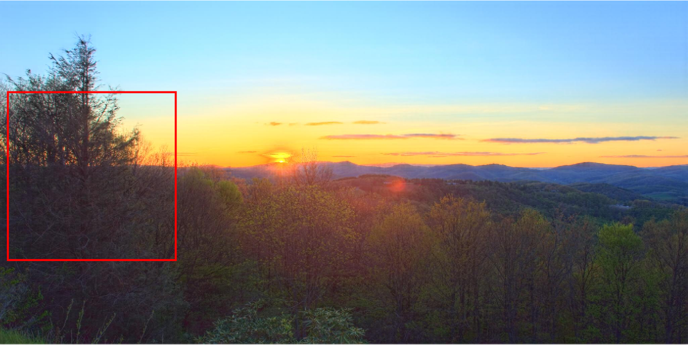}
		\label{fig:ourmodell1}
	\end{subfigure}

	\begin{subfigure}[t]{0.235\textwidth}
		\includegraphics[width=\linewidth,height=.7\linewidth]{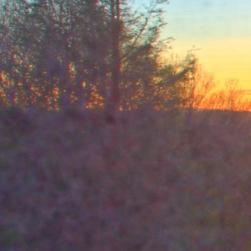}
	\end{subfigure}
	\begin{subfigure}[t]{0.235\textwidth}
		\includegraphics[width=\linewidth,height=.7\linewidth]{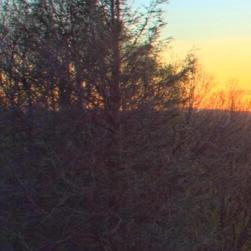}
	\end{subfigure}	                                                                                \caption{\small Comparison between CNN (encoder-decoder) with $\mathcal{L}_1$-loss and DeepTMO (single-scale). Inlets in row 2 show that DeepTMO yields sharp and high resolution output, whereas the CNN results in blurred outputs. }
	\label{fig:OurvsL1}
	
\end{figure}

\begin{figure}[h]
	\begin{subfigure}[t]{0.23\textwidth}
		\includegraphics[width=\linewidth]{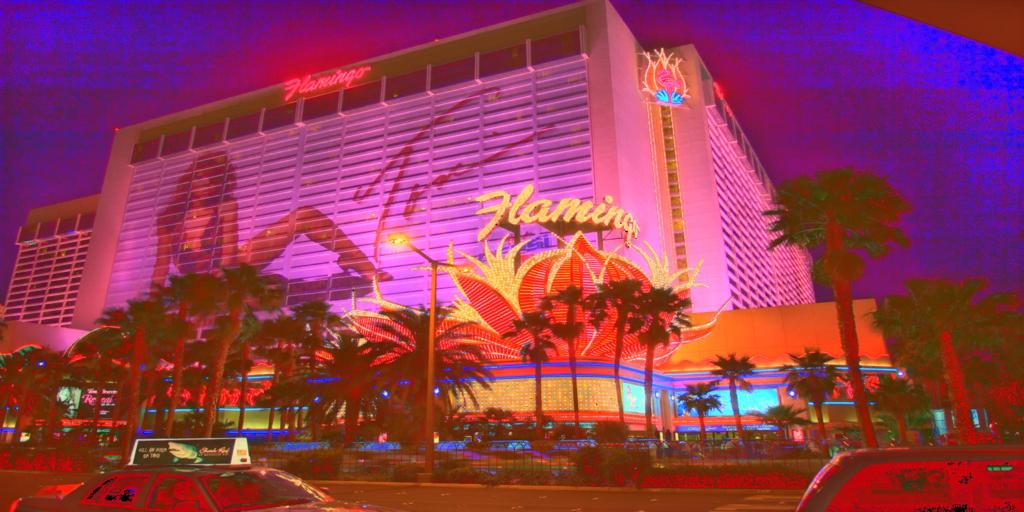}
		\caption{$\mathcal{L}_{prp}$-loss}
		\label{fig:lploss}
	\end{subfigure}
	\begin{subfigure}[t]{0.23\textwidth}
		\includegraphics[width=\linewidth]{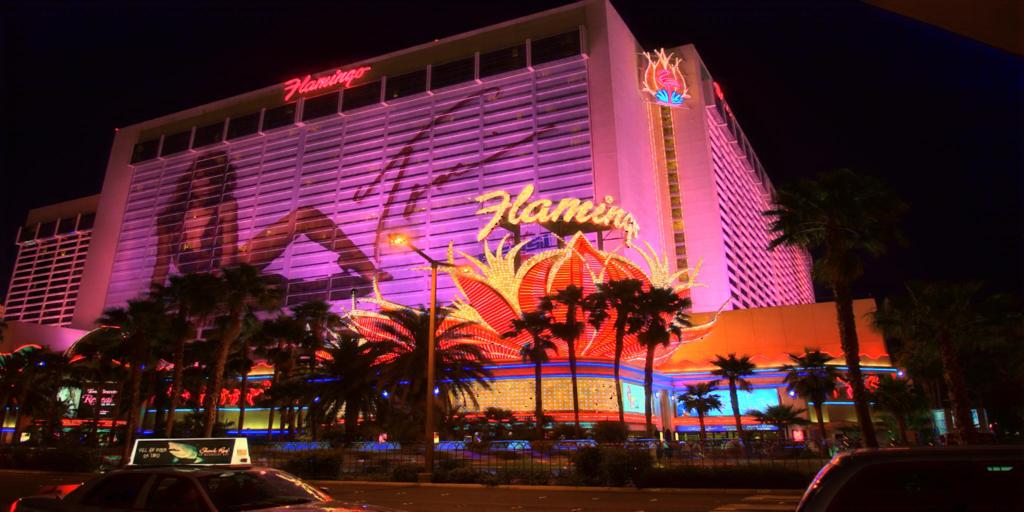}
		\caption{Deep-TMO (Single-Scale)}
		\label{fig:ourmodellp}
	\end{subfigure}
	\caption{\small Comparison between CNN (encoder-decoder) with $\mathcal{L}_{prp}$-loss (perceptual) and DeepTMO (single-scale). 
	}
	\label{fig:OurvsLp}
\end{figure}

\paragraph{Motivation for GANs}
To achieve the desired TMO, one solution is to use a simple $\mathcal{L}_1$ or perceptual ($\mathcal{L}_{prp}$) loss function~\cite{johnson2016perceptual} with an encoder-decoder architecture as utilized in the past by various inverse-TMOs for generating HDR scenes from single-exposure \cite{eilertsen2017hdr} or multi-exposure \cite{deepreverse} LDR images. However, such naive loss functions suffer from either overall spatial blurring (evident in $\mathcal{L}_1$ loss in Fig.~\ref{fig:OurvsL1}) or over-compression of contrast (evident in $\mathcal{L}_{prp}$ loss in Fig.~\ref{fig:OurvsLp}). This is mainly because a CNN architecture learns a mapping from all possible dynamic range values available in the wide variability of training-set scenes to a range [0,255]. Thus, the trained model effectively predicts a fairly mean luminance value for most of the pixels in output images to minimize the overall loss function. Another simple idea could be to use TMQI directly as loss function. However, due to the mathematical design of TMQI's naturalness component and characteristic discontinuity, TMQI cannot be directly used a loss function for back-propagation in a DL framework. In fact, the alternate methodology proposed by authors in~\cite{Tmqi2}, which optimizes a given TMO using TMQI, is also impossible to be imbibed into an end-to-end DL pipeline, as it treats both SSIM and naturalness separately using two different optimization strategies.

Given the goal of our TMO, designing an effective cost function manually for catering to wide variability of tone-mapping characteristics under different scenic-content is quite a complex task. An alternate solution could be to \textit{learn} such a loss function. The use of GAN is an apt choice here, as it learns an adversarial loss function by itself (the loss being the discriminator network), that encapsulates all the desired features for an ideal TMO encoded in the underlying training data, thereby eradicating the need of manually designing a loss function. An added advantage of GAN is that it facilitate to obtain perceptually superior tone-mapped solutions residing in the subspace of natural images as compared to reproducing closer to mean valued or blurred outputs in case of ordinary $\mathcal{L}_1$ / $\mathcal{L}_{prp}$ loss functions.
 
Aiming an artifact-free high-resolution tone-mapped output, we begin investigating the choice of architecture from single-scale to its multi-scale variant for both generator and discriminator in the following sections.

\begin{figure}[h]
	\includegraphics[width=\linewidth]{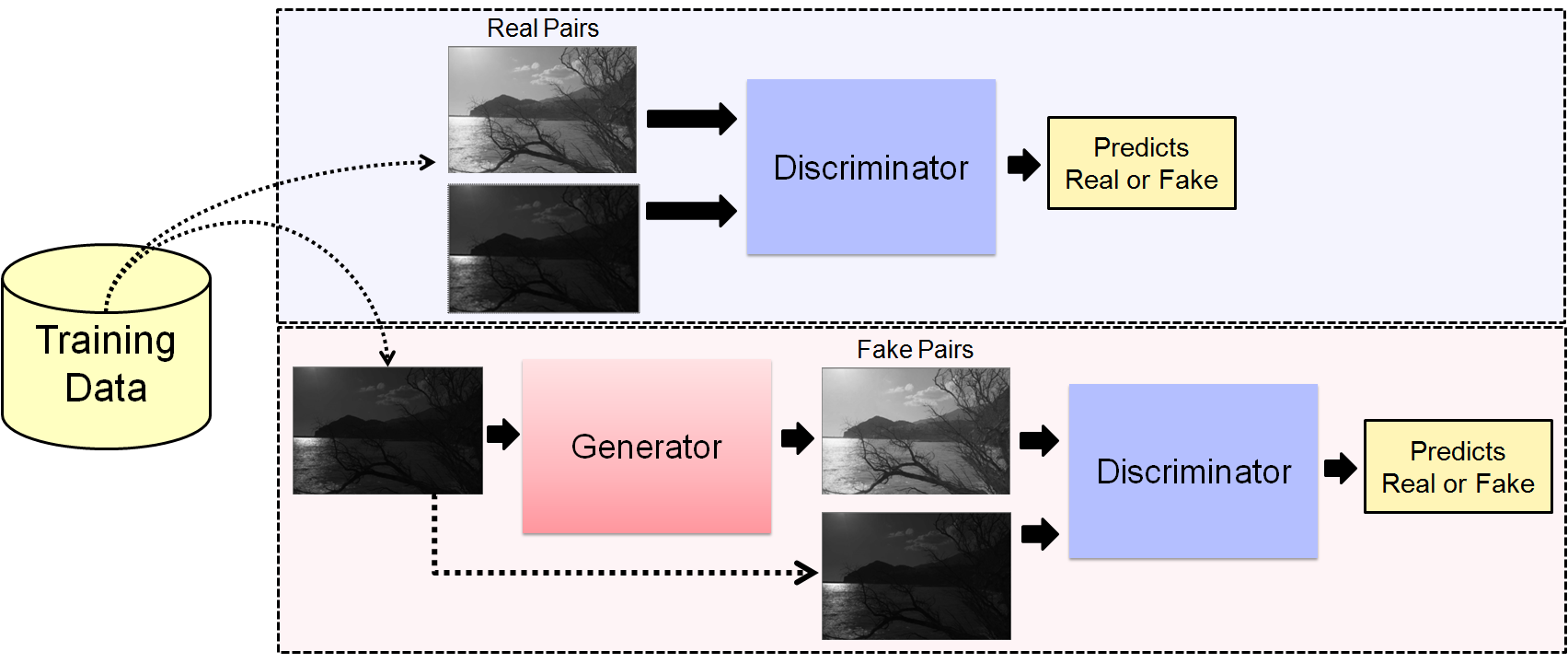}
	\caption{\small DeepTMO Training Pipeline. 
		}
	\label{fig:overview}
\end{figure}

\subsection{DeepTMO (Single-Scale)}   
Fig.~\ref{fig:overview} depicts an overview of our training algorithm. For our DeepTMO model, we basically employ a cGAN framework~\cite{cgan} which implicitly learns a mapping from an observed HDR image $x$ to a tone mapped LDR image $y$, given as: $ G : {x} \longrightarrow y$. The architecture is composed of two fundamental building blocks namely a discriminator $(D)$ and a generator $(G)$. 

The input to $G$ consists of an $H \times W \times C$ size HDR image normalized between $[0,1]$. We consider $C = 1$ \ie only the luminance channel is given as an input. Its output is a tone-mapped image (top row of fake pair in Fig.~\ref{fig:overview}) of same size as the input.  
$D$ on the other hand, takes luminance channels of HDR and tone mapped LDR images as input pairs, and predicts whether they are real tone-mapped images or fake. It is trained in a supervised fashion, by employing a training dataset of input HDR and their corresponding \textit{target} tone-mapped images (real-pair in Fig.~\ref{fig:overview}). We detail the complete methodology to build our target dataset in Section~\ref{sec:buildingdataset}. 
An additional advantage of conditioning on an HDR input is that it empowers $D$ to have some pre-information to make better reasoning for distinguishing between a real or fake tone mapped images, thus accelerating its training.

Next, we discuss the architectures for single-scale generator (Single-G) and single-scale discriminator (Single-D) which are our adaptations from past studies~\cite{johnson2016perceptual,cyclegan} which show impressive results for style transfer and super-resolution tasks on LDR images. Further on, in the subsequent sections, we will reason as to why opting for their multi-scale versions aids in further refining the results. 

\begin{figure*}[h]
	\begin{subfigure}[t]{0.4\textwidth}
		\includegraphics[width=\linewidth]{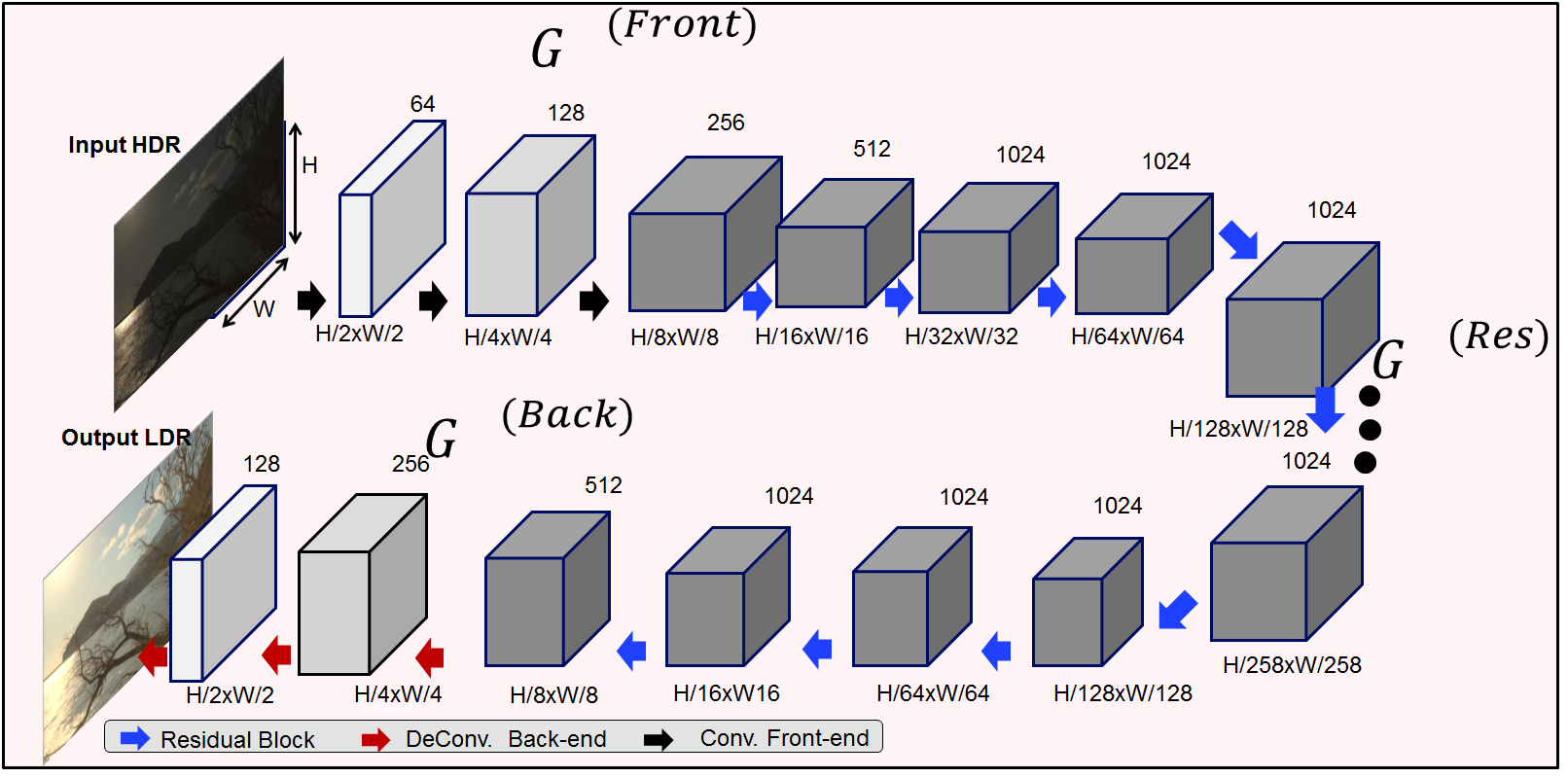}
		\caption{Generator (Single Scale).}
		\label{fig:generator}
	\end{subfigure}\hspace{0.5cm}
	\begin{subfigure}[t]{0.3\textwidth}
		\includegraphics[width=\linewidth]{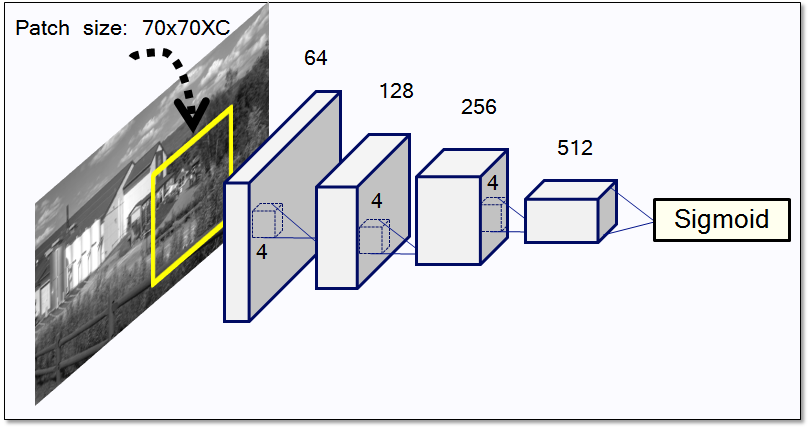}
		\caption{Discriminator (Single Scale).}
		\label{fig:discriminator}
	\end{subfigure}\hspace{0.5cm}
	\begin{subfigure}[t]{0.2\textwidth}
		\includegraphics[width=\linewidth]{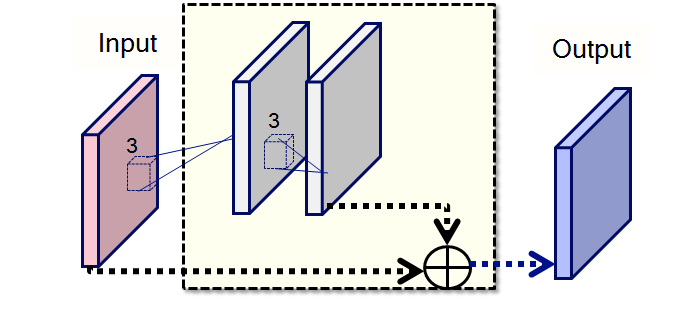}
		\caption{Residual Blocks}
		\label{fig:resblock}
	\end{subfigure}	
	\caption{\small DeepTMO (single-scale) generator and discriminator architecture. The generator in (a) is an encoder-decoder architecture. 
	Residual blocks in (c) consist of two sequential convolution layers applied to the input, producing a residual correction. 
	Discriminator in (b) consists of a patchGAN \cite{pix2pix, li2016precomputed, ledig2016photo} architecture which is applied patch wise on the concatenated the input HDR and tone mapped LDR pairs. 
	More details in Supplementary.}
	\label{fig:gendisc} 
\end{figure*}

\paragraph{Generator Architecture (Single-G)}
The Single-G architecture is an encoder-decoder architecture as shown in Fig.~\ref{fig:generator}. Overall, it consists of a sequence of 3 components: the convolution front end $G^{(Front)}$, a set of residual blocks $G^{(Res)}$ and the deconvolution back end $G^{(Back)}$. $G^{(Front)}$ consists of 4 different convolution layers which perform a subsequent down-sampling operation on their respective inputs. $G^{(Res)}$ is composed of 9 different residual blocks each having 2 convolution layers, while $G^{(Back)}$ consists of 4 convolution layers each of which up-samples its input by a factor of 2. During the down-sampling, $G^{(Front)}$ compresses the input HDR, thus keeping the most relevant information. $G^{(Res)}$ than applies multiple residual corrections to convert the compressed representation of input HDR to one that of its target LDR counterpart.
 
Finally, $G^{(Back)}$ yields a full size LDR output from this compressed representation through the up-sampling operation. 
	
\paragraph{Discriminator Architecture (Single-D)}
The Single-D architecture resembles a $70 \times 70$ PatchGAN \cite{pix2pix, li2016precomputed, ledig2016photo} model, which aims to predict whether each $70 \times 70$ overlapping image patch is real or fake, as shown in Fig.~\ref{fig:discriminator}. The main motivation of choosing a PatchGAN discriminator over a full-image size discriminator is that it contains much less parameters allowing it to be easily used for any-size images in a fully convolutional manner. This is pertinent for our problem setting where we involve very high resolution images. An added advantage of a PatchGAN discriminator is that while working on patches, it also models the high-frequency information by simply restricting its focus upon the structure in local image regions. The Single-D is run across the entire image, and all the responses over various patches are averaged out to yield the final prediction for the image. Note that the input to $D$ is a concatenation of the HDR and its corresponding LDR image.  

Although the Single-G and Single-D architecture yields high-quality reconstructions at a global level, yet it results in noisy artifacts over some specific areas such as bright light sources as shown in Fig.~\ref{fig:SSDSSG}. In a way, it necessitates modifying both single-scale versions of $G$ and $D$ to cater not only to coarser information, but at the same time, paying attention to finer level details, thus resulting in a much more refined tone-mapped output.

\subsection{DeepTMO (Multi-Scale)}
While generating high resolution tone-mapped images, it is quite evident now that we need to pay attention towards low-level minute details as well as high-level semantic information. To this end, motivated from~\cite{pix2pixhd}, we alter the existing DeepTMO (single-scale) model, gradually incorporating step-by-step a multi-scale discriminator (Multi-D) and a multi-scale generator (Multi-G) in the algorithmic pipeline. Different from~\cite{pix2pixhd}, our adaptation (a) utilizes a 2-scaled discriminator, (b) incorporates a different normalization layer in the beginning given by $\frac{(x-x_{min})}{(x_{max}-x_{min})}$, scaling pixels between [0,1] with a high 32-bit floating point precision, (c) inputs specifically a single luminance channel input with 32-bit pixel-depth linear HDR values.

In the following, we detail the multi-scale versions of $G$ and $D$. We showcase the impact through step-wise substitution of the Single-D with its Multi-D variant, and then the Single-G as well with its Multi-G counterpart.
\begin{figure}[h]
	\begin{subfigure}[t]{0.23\textwidth}
		\includegraphics[width=\linewidth]{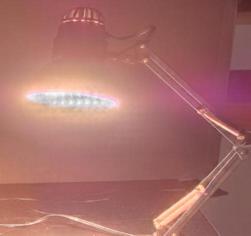}
		\caption{Single-D \& Single-G }
		\label{fig:SSDSSG}
	\end{subfigure}
	\begin{subfigure}[t]{0.23\textwidth}
		\includegraphics[width=\linewidth]{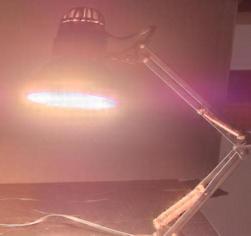}
			\caption{Multi-D \& Single-G}
		\label{fig:MSDSSG}
	\end{subfigure}
	
	\begin{subfigure}[t]{0.23\textwidth}
		\includegraphics[width=\linewidth]{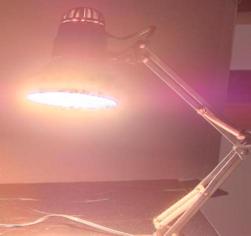}
		\caption{Single-D \& Multi-G }
		\label{fig:SSDMSG}
	\end{subfigure}
	\begin{subfigure}[t]{0.23\textwidth}
		\includegraphics[width=\linewidth]{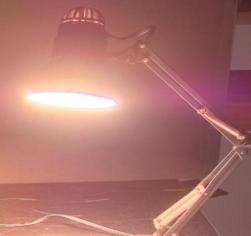}
		\caption{Multi-D \& Multi-G}
		\label{fig:MSDMSG}
	\end{subfigure}
	
	\caption{\small Impact of Multi-scale Discriminator and Generator.}
	\label{fig:MSDMSGimpact}
\end{figure}

\paragraph{Multi-D}
Correctly classifying a high-resolution tone-mapped output as real or fake is quite challenging for Single-D. Even though an additional loss term effectively removes noisy artifacts at a global scale in the image (illustrated later in Section~\ref{sec:tmof}), we still witness repetitive patterns in specific localized regions while using Single-D (for \eg seen around high illumination sources like inside/outside the ring of table lamp in Fig.~\ref{fig:SSDSSG} and on the ring of the lamp in Fig~\ref{fig:SSDMSG}). One easy way to tackle this problem is by focusing the discriminator's attention to a larger receptive field which is possible either through a deeper network or larger convolution kernels. However, it would in-turn demand a higher memory-bandwidth, which is already a constraint for training high-resolution HDR images. Thus, we basically retain the same network architecture for the discriminator as used previously, but rather apply it on two different scales of input \ie the original and the $2 \times$ down-sampled version, calling the two discriminators $D_{o}$ and $D_{d}$ respectively. 

Both $D_{o}$ and $D_{d}$ are trained together to discriminate between real and synthetically generated images. $D_{d}$, by working on a coarser scale, focuses on a larger area of interest in patches throughout the image. This feature subsequently aids $G$ to generate more globally consistent patch-level details in the image. $D_{o}$ on the other hand, operating at a much finer scale than $D_{d}$, aids in highlighting more precise finer nuances in patches, thus enforcing $G$ to pay attention towards very minute details too at the time of generation. Thus, by introducing a Multi-D, the noisy patterns observed in Single-D, are suppressed to a large extent (for \eg as seen in Fig.~\ref{fig:SSDSSG} and Fig.~\ref{fig:MSDSSG}). However, we still witness minor traces of these artifacts due to Single-G's very own limitations, thus compelling us to switch to Multi-G. Contrary to Single-G, Multi-G reproduces outputs taking notice of both coarser and finer scales. Thus, the resultant output, having information over both scales, yields a more globally consistent and locally refined artifact-free image (for \eg as seen in Fig.~\ref{fig:MSDSSG} and Fig.~\ref{fig:MSDMSG}.

\begin{figure}
	\includegraphics[width=\linewidth]{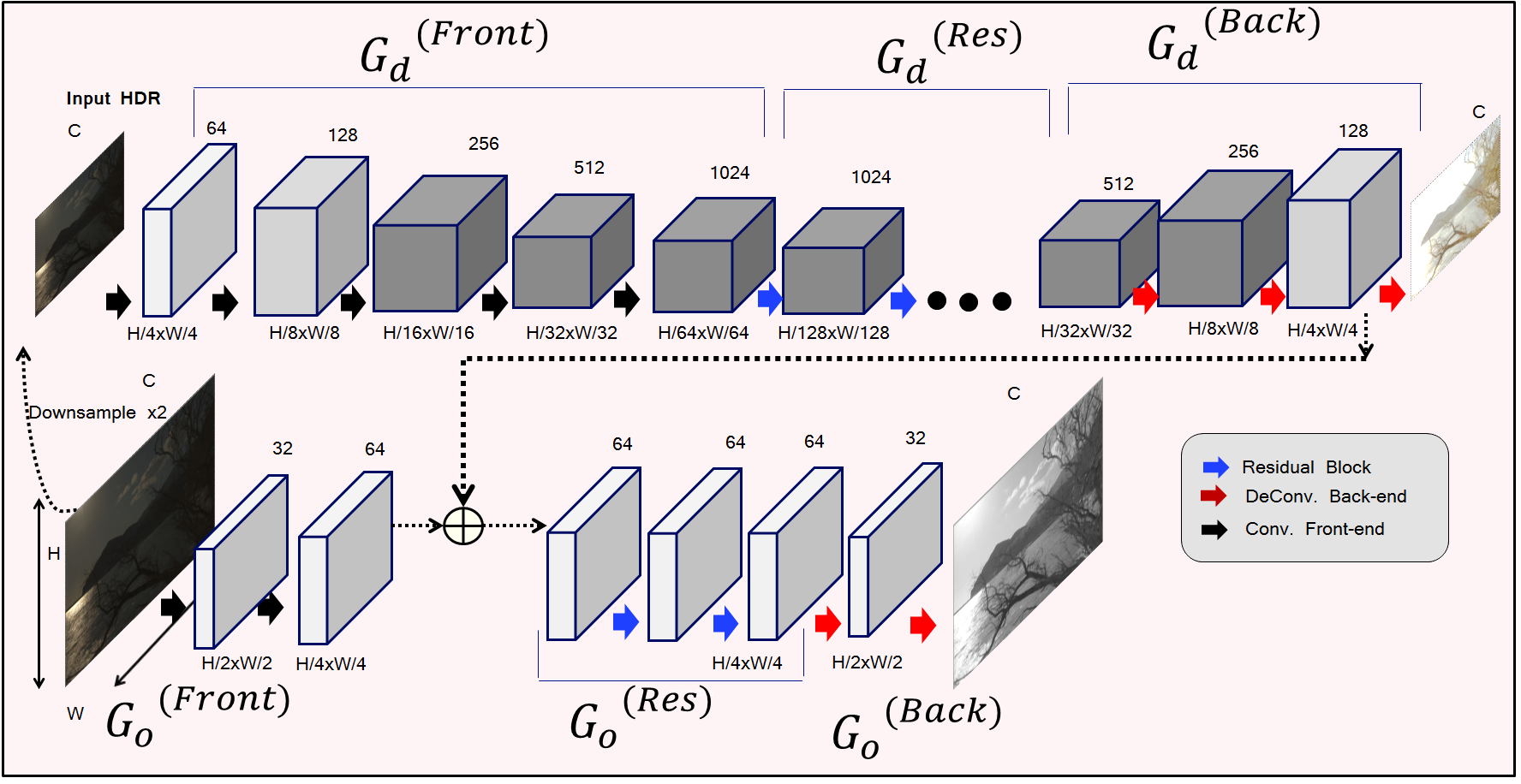}
	\caption{\small DeepTMO multi-scale generator architecture. While the finer generator $G_{o}$ has the original image as its input, the input to $G_{d}$ is a $2 \times$ down-sampled version.}
	\label{fig:deeptmoh}
\end{figure}

\paragraph{Multi-G}
Fig.~\ref{fig:deeptmoh} illustrates the design of Multi-G. It mainly comprises of two sub-architectures, a global down-sampled network $G_{d}$ and a global original network $G_{o}$. The architecture for $G_{d}$ is similar to Single-G with the components, convolutional front-end, set of residual blocks and convolutional back-end being represented as: $G_{d}^{(Front)}, G_{d}^{(Res)}, G_{d}^{(Back)}$, respectively. 
$G_{o}$ is also similarly composed of three components given by: $G_{o}^{(Front)}$, $G_{o}^{(Res)}$ and $G_{o}^{(Back)}$. 
	
As illustrated in Fig.~\ref{fig:deeptmoh}, at the time of inference, while the input to $G_{o}$ is a high resolution HDR image ($2048 \times 1024$), $G_{d}$ receives a $2 \times$ down sampled version of the same input. $G_{o}$ effectively makes tone-mapped predictions, paying attention to local fine-grained details (due to its limited receptive field on a high resolution HDR input). At the same time, it also inputs from $G_{d}$, a coarser prediction (as its receptive field has a much broader view). Thus, the final generated output from $G_{o}^{(Back)}$ encompasses local low-level information and global structured details together in the same tone-mapped output. Hence, what we finally obtain is a much more structurally preserved and minutely refined output which is free from local noisy-artifacts, as seen in Fig.~\ref{fig:MSDMSG}. 

To summarize, we showcase 4 different cGAN designs where the:
\begin{enumerate}
	\item Single-D \& Single-G architecture encounters noisy patterns due to not paying attention to finer-level details.
	\item Multi-D \& Single-G architecture is able to suppress patterns to some extent as observed in the previous case. This is mainly due to limited generalization capabilities of Single-G. 
	\item Single-D \& Multi-G architecture removes patterns throughout the image, however some very localized regions still face artifacts due to the limited capacity of Single-D.  
	\item Multi-D \& Multi-G architecture finally yields superior quality artifact-free images. 
\end{enumerate}

\subsection{Tone Mapping Objective Function} \label{sec:tmof} 
The ultimate goal of $G$ is to convert high resolution HDR inputs to tone mapped LDR images, while $D$ aims to distinguish real tone-mapped images from the ones synthesized by $G$. We train both the $G$ and $D$ architectures in a fully supervised setting. For training, we give a set of pairs of corresponding images $\{( x_{i} , y_{i} )\}$, where $x_{i}$ is the luminance channel of the HDR input image while $y_{i}$ is the luminance channel output of the corresponding tone-mapped LDR image. Next, we elaborate upon the objective function to train our DeepTMO (both single-scale and multi-scale). 

The basic principle behind cGAN~\cite{cgan} is to model the conditional distribution of real tone-mapped images given an input HDR via the following objective:
\begin{equation}
\mathcal{L}_{cGAN}(G, D) = E_{x,y}[\log D(x,y)]+ E_{x}[\log (1 - D(x, G(x)))],
\end{equation}

where G and D compete with each other; G trying to minimize this objective against its adversary D, which tries to maximize it, i.e.
$G^{\ast} = \arg \min_{G} \max_{D} \mathcal{L}_{cGAN} (G, D)$.

Since the Multi-D architecture consists of $D_{o}$ and $D_{d}$, our objective for the same is:

\begin{equation}
G^{\ast} = \arg \min_{G} \max_{D_{o},D_{d}} \sum_{s = o,d} \mathcal{L}_{cGAN} (G, D_{s}) 
\end{equation}\label{eq:Gobjective2}
We append to the existing cGAN loss, an additional regularization term in the form of a feature matching (FM) loss $\mathcal{L}_{FM}(G,D_{s})$ (similar to perceptual loss \cite{percept1,percept2}), given by:
\begin{equation}
\mathcal{L}_{FM}(G,D_{s}) = {E}_{(x,y)} \sum_{i=1}^{M} \frac{1}{U_{i}} [|| D_{s}^{(i)}(x, y) -  D_{s}^{i}(x, G(x))||_{1}],
\end{equation}

where $D_s^{i}$ is the $i^{th}$ layer feature extractor of $D_{s}$ (from input to the $i^{th}$ layer of $D_{s}$), $M$ is the total number of layers and $U_{i}$ denotes the number of elements in each layer. In short, we extract features from each individual $D$ layer and match these intermediate representations over real and generated images. Additionally, we append a perceptual loss $L_{prp}$ as used in \cite{johnson2016perceptual}, which constitutes of features computed from each individual layer of a pre-trained 19-layer VGG network ~\cite{simonyan2014very} given by:
$\mathcal{L}_{L_{prp}} (G) = \sum_{i=1}^{N} \frac{1}{V_{i}} [||F^{(i)} (y) - F^{(i)}(G(x))||_{1}]$

where $F^{(i)}$ denotes the $i^{th}$ layer with $V_{i}$ elements of the VGG network. The VGG network had been pre-trained for large scale image classification task over the Imagenet dataset ~\cite{russakovsky2015imagenet}. Henceforth, our final objective function for a DeepTMO can be written as:

\begin{equation}
\begin{split}
G^{\ast} = \arg \min_{G} \max_{D_{o},D{d}} \sum_{s = o,d} \mathcal{L}_{cGAN} (G, D_s) + \\
\beta \sum_{s = o,d} \mathcal{L}_{FM} (G,D_{s})
+ \gamma \mathcal{L}_{L_{prp}} (G)
\end{split}	
\end{equation}

$\beta$ and $\gamma$ controls the importance of $\mathcal{L}_{FM}$  and $\mathcal{L}_{L_{prp}}$ with respect to $\mathcal{L}_{cGAN}$ and both are set to 10. We illustrate the impact of both these terms in the following paragraph. 

\begin{figure}
	\begin{subfigure}[t]{0.23\textwidth}
		\includegraphics[width=\linewidth]{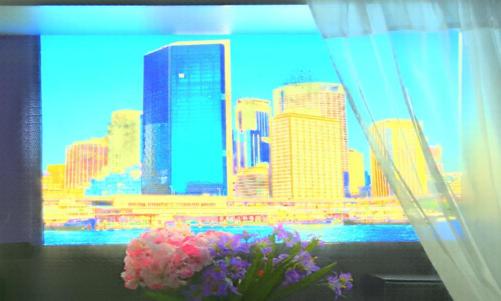}
		\caption{without VGG and FM Loss}
		\label{fig:woVGGandFM}
	\end{subfigure}
	\begin{subfigure}[t]{0.23\textwidth}
		\includegraphics[width=\linewidth]{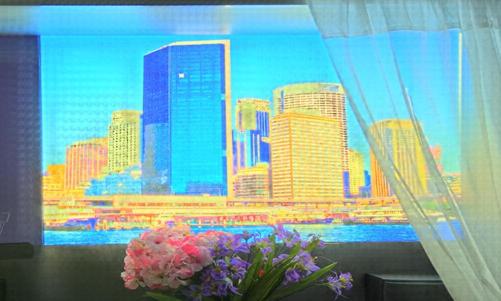}
		\caption{without VGG loss}
		\label{fig:woVGG}
	\end{subfigure}
	
	\begin{subfigure}[t]{0.23\textwidth}
		\includegraphics[width=\linewidth]{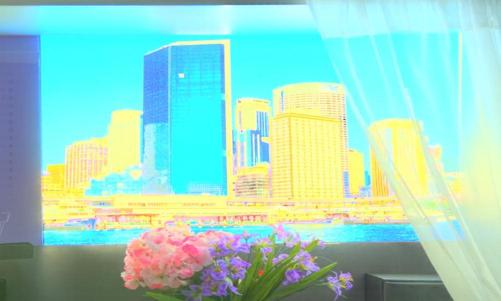}
		\caption{without FM Loss}
		\label{fig:woFM}
	\end{subfigure}
	\begin{subfigure}[t]{0.23\textwidth}
		\includegraphics[width=\linewidth]{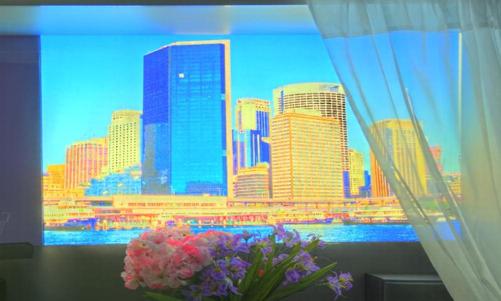}
		\caption{with VGG and FM loss}
		\label{fig:withVGGandFM}
	\end{subfigure}
	
	\caption{\small DeepTMO (single-scale) with/without FM and VGG Loss.}
	\label{fig:w/oVGGandFM}
\end{figure}

\paragraph{Impact of Feature Matching and Perceptual Loss term}

Both $\mathcal{L}_{FM}$  and $\mathcal{L}_{L_{prp}}$ loss terms act as guidance to the adversarial loss function preserving overall natural image statistics and training without both these terms results in inferior quality throughout the image. The VGG-term primarily checks for global noisy repetitive patterns in the image and helps in suppressing them. While being applied on the full generated image, the VGG network captures both low-level image characteristics (\eg fine edges, blobs, colors etc,) and the high level semantic information through its beginning-level and later-stage network layers, respectively. Based upon these features, VGG effectively detects the corresponding artifacts as a shortcoming in the overall perceptual quality of the generated scene and hence guides to rectify them; thereby yielding a more natural image. For \eg the removal of noisy can be visualized by looking simultaneously at Fig.~\ref{fig:woVGG} and~\ref{fig:withVGGandFM}. The FM loss term on the other hand, caters to more localized quality 
details like keeping a watch on illumination conditions in each sub-region. For \eg it effectively tones-down over-exposed regions of windows in the building as can be seen in Fig.~\ref{fig:woFM} and~\ref{fig:withVGGandFM}. This is ideally done by utilizing various feature layers of $D$, which are trained by focusing upon $70\times70$ localized image patches. Together both (VGG and FM) loss terms help in yielding a high quality overall contrast and local finer-details preserved output image (as seen in Fig.~\ref{fig:withVGGandFM}).

\subsection{Network Insight} \label{sec:nwi}
Every component in the network plays an indispensable role in the overall tone-mapping. Starting from the convolutional front ends $G_{d}^{(Front)}$ and $G_{o}^{(Front)}$, both of which are applied directly on the linear HDR input, compress its tone and transform it to an encoded representation in an HDR space. While the convolutional layers play a critical role in down-sampling the spatial resolution by deriving meaningful feature layers using its learnt filters, the Instance Norm and activation functions (following each conv-layer) help in compressing the dynamic range of each pixel intensity. Next, the residual layers $G_{d}^{(Res)}$ and $G_{o}^{(Res)}$ can be understood as functions that map the current encoded information in HDR space to one in the LDR space. This is essentially accomplished by adding a residual information to the current compressed form of HDR input. Finally, the $G_{d}^{(Back)}$ and $G_{o}^{(Back)}$ are applied to this encoded representation in the LDR space in order to transform it into a rich and visually pleasing LDR output. While the transposed convolution pay special attention to spatial upsampling, the activation functions maintains a compressed tone which is perceptually `the most' appealing for a given scene.

\begin{table}[h]
	\caption{\footnotesize Abbreviations}\label{tab:tmqi}
	\centering
	\begin{tabular}{c|c}
		\centering
		Symbols & Meaning \\ 
		\hline
		$G_o$, $G_d$ & Generator Original, Downsampled scale  \\
		$D_o$, $D_d$ & Discriminator  Original, Downsampled scale \\
		$\mathcal{L}_{cGAN}$ & Adversarial Loss \\
		$\mathcal{L}_{FM}$ & Feature Matching Loss \\
		$\mathcal{L}_{prp}$ & Perceptual Loss\\
		$\mathcal{L}_{1}$ & $L_1$ Absolute Loss\\
		$H, W, C$ & Height, Width and Channel\\
		$\beta, \gamma$ & control parameters\\
		\hline
	\end{tabular}
\end{table}
	\section{Building the HDR Dataset}\label{sec:buildingdataset}
In order to design a deep CNN based TMO, it is essential to obtain a large-scale dataset with a wide diversity of real-world scenes that are captured using a variety of cameras. To this end, we gather the available HDR datasets. For training the network, a total of 698 images are collected from various different sources, listed in the supplementary material. From the HDR video dataset sources, we select the frames manually so that no two chosen HDR images are similar. All these HDR images have been captured from diverse sources which is beneficial for our objective \ie learning a TMO catering a wide variety of real-world scenes.

To further strengthen the training, we applied several data augmentation techniques such as random cropping and flipping, which are discussed briefly in section \ref{sec:implem}. We considered 105 images from the~\cite{database2} for testing purposes.

\subsection{Target Tone Mapped Images}\label{sec:tmqi}
Selecting a `target' tone mapped image for a given HDR scene is a crucial step in training the DeepTMO. Although several subjective studies~\cite{Banterle} built on different hypotheses have attempted to answer this question, yet they have been conducted only for very small databases of sizes upto 15-20 scenes. Such subjectively evaluated databases are limited in number and cannot be effectively used as training dataset for our DeepTMO model. Additionally, these databases have been evaluated under varying evaluation setups \ie by using different sets of TMOs and reference or no-reference settings. Hence, similar to~\cite{patel}, we resorted to a widely used objective-metric known as TMQI~\cite{tmqi} to ensure a fixed target selection criterion for our problem.
 
As discussed in Section 2, literature of TMOs is quite extensive and practically difficult to span. Therefore, to find the target tone mapped image for each training HDR scene, we selected 13 classical TMOs:~\cite{ward97,patt02,ashikhmin02,durand02,tumblin99,drago03,schlick94,reinhard02,fattal02,chiu96,mantiuk06} and gamma and log mappings~\cite{Banterle}. The selection of these tone mappings is inspired from the subjective evaluation studies~\cite{ledda2005evaluation,tmqi,Tmqi2,cadik08} which highlight the distinctive characteristics of mapping functions, which we aim to inculcate into the learning of our DeepTMO model.

For each HDR scene, we initially rank the obtained tone-mapped outputs from all the 13 TMOs using the TMQI metric. Then, the best scoring tone mapped output is selected as the `target' for the corresponding HDR scene. Since tuning the parameters of 13 considered TMOs is a daunting task for a large set of training images, we used their default parameter settings throughout this paper. Though we acknowledge that fine-tuning TMO parameters can further boost overall performance, the process however, is almost impractical considering the large amount of training images and the vast parameter-space of the TMOs.

	\section{Training and Implementation Details}  
DeepTMO training paradigm is inspired by the conventional GANs approach, where alternate stochastic gradient descent (SGD) steps are taken for $D$ followed by the $G$. We specifically utilize Least Square GANs (LSGANs), which have proven to yield~\cite{LsGANs} a much more stable learning process compared to regular GANs. For the multi-scale architecture, we first train $G_{d}$ separately, and then fine tune both $G_{d}$ and $G_{o}$ (after freezing the weights of $G_{d}$ for the first 20 epochs). For both $D$ and $G$, all the weights corresponding to convolution layers are initialized using zero mean Gaussian noise with a standard deviation of $0.02$, while the biases are set to $0$. 
\begin{figure}[h]
	\begin{subfigure}[t]{0.24\textwidth}
		\includegraphics[width=\linewidth]{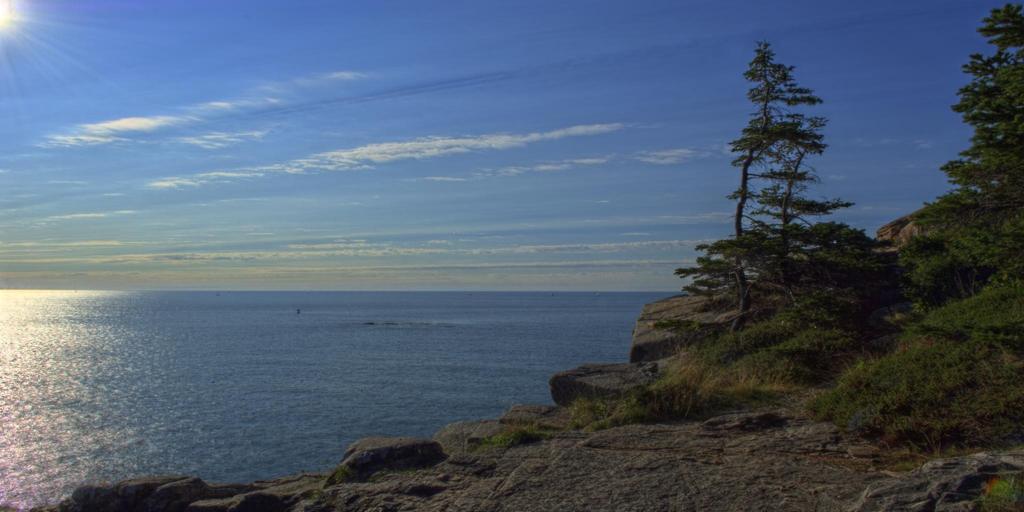}
		\caption{With Instance Norm}
		\label{fig:IN2}
	\end{subfigure}
	\begin{subfigure}[t]{0.24\textwidth}
		\includegraphics[width=\linewidth]{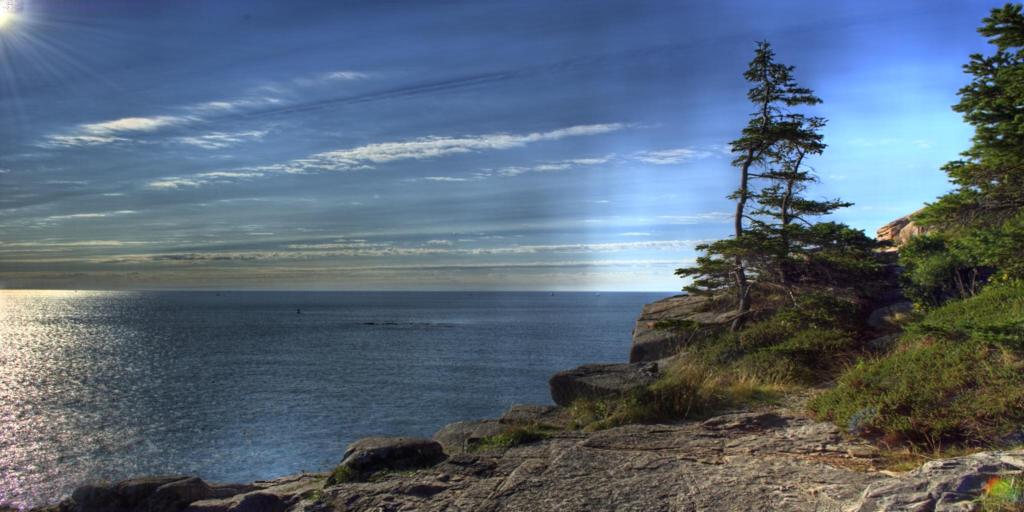}
		\caption{With Batch Norm}
		\label{fig:BN2}
	\end{subfigure}
	
	\caption{\small Batch Normalization vs. Instance Normalization.}
	\label{fig:BNIN}
\end{figure}

\subsubsection{Instance Vs. Batch Norm}
We use instance normalization~\cite{instancenorm}, which is equivalent to applying batch normalization~\cite{batchnorm} using a batch size equal to 1. 

The efficacy of the instance-norm is showcased in Fig.~\ref{fig:BNIN}, where applying the plain batch-norm yields non-uniformity in luminance compression. While the instance normalization is trained to learn mean and standard deviation over a single-scene for the purpose of normalizing, the batch-norm learns over a full batch of input images. Thus, its mean and standard deviation is computed spatially for each pixel from a much wider range of high dynamic luminance values over the entire batch leading to uneven normalization. 

Absence of batch-norm/instance-norm prevents the $G$/$D$ to train properly and results in poor generation quality, thus necessitating the need for a normalization layer. All the instance normalization layers are initialized using Gaussian noise with mean 1 and $0.02$ standard deviation. 

\subsubsection{Implementation}\label{sec:implem}
All training experiments are performed using the Pytorch \cite{pytorch} deep learning library with mini-batch SGD, where the batch size is set to $4$. For multi-scale, we use batch-size 1 due to limited GPU memory. We utilize an ADAM solver \cite{adam} with initial learning rate fixed at $2 \times 10^{-4}$ for the first 100 epochs and then, allowed to decay to 0.0 linearly, until the final epoch. Momentum term $\beta_{1}$ is fixed at $0.5$ for all the epochs. Hyper-parameters have been set to their default values and aren’t manipulated much due to GANs training complexity. We also employ random jitters by first resizing the original image to $700 \times 1100$, and then randomly cropping to size $512 \times 512$. For multi-scale, we resize to $1400 \times 2200$ and crop to size $1024 \times 1024$. All our networks are trained from scratch. 

For all the other handcrafted TMOs, we used the MATLAB-based HDR Toolbox~\cite{Banterle} and Luminance HDR software~\footnote{http://qtpfsgui.sourceforge.net/}. For each TMO, we enabled the default parametric setting as suggested by the respective authors. Training is done using a 12 Gb NVIDIA Titan-X GPU on a Intel Xeon e7 core i7 machine for 1000 epochs and takes a week. 

	\section{Results and Evaluation} 
In this section, we present the potential of our DeepTMO on a wide range of HDR scenes, containing both indoor and outdoor, human and structures, as well as day and night views. We compare our results with the best subjective outputs obtained from wide range of tone mapping  methods~\cite{drago03,mantiuk06,durand02,schlick94,chiu96,patt02,ashikhmin02,reinhard02} on $105$ images of test dataset~\cite{database2}, both qualitatively and quantitatively. In addition, we briefly discuss the specific characteristics of the proposed model, including their adaptation to content or sharpness in displaying high-resolution tone mapped outputs. Finally, we present a subjective evaluation study to access the perceived quality of the output. The size for each input image is kept fixed at $1024\times2048$. 

Note that test scenes are different from the training set and are not seen by our model while training. Full size images and some additional results can be found in the supplementary material for better visual quality.
\begin{figure*}
	\captionsetup[subfigure]{labelformat=empty} 
	\begin{subfigure}[b]{.01\textwidth}
		\raisebox{.3in}{\rotatebox[origin=t]{90}{\footnotesize Scene-1}}   
	\end{subfigure}	
	\begin{subfigure}[b]{.243\textwidth}
		\caption{HDR Linear}
		\includegraphics[width=\linewidth]{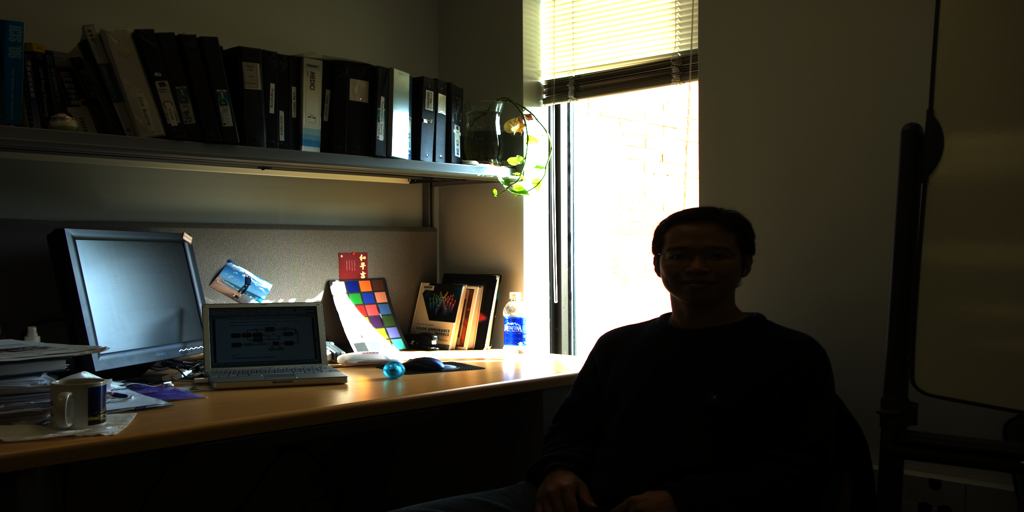}
	\end{subfigure} 
	\begin{subfigure}[b]{.243\textwidth}
		\caption{DeepTMO (.80)}
		\includegraphics[width=\linewidth]{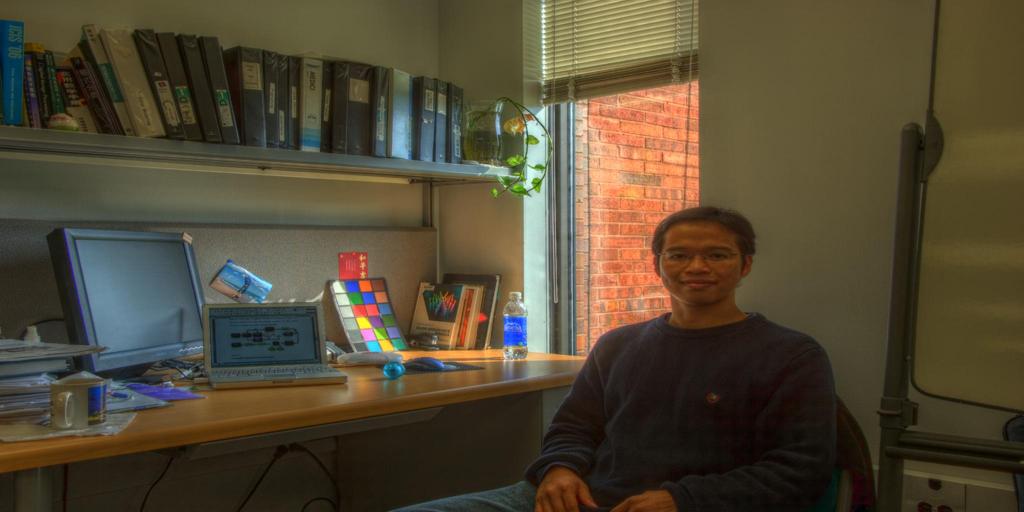}
	\end{subfigure} 	
	\begin{subfigure}[b]{.243\textwidth}
		\caption{MantiukTMO (0.81)}
		\includegraphics[width=\linewidth]{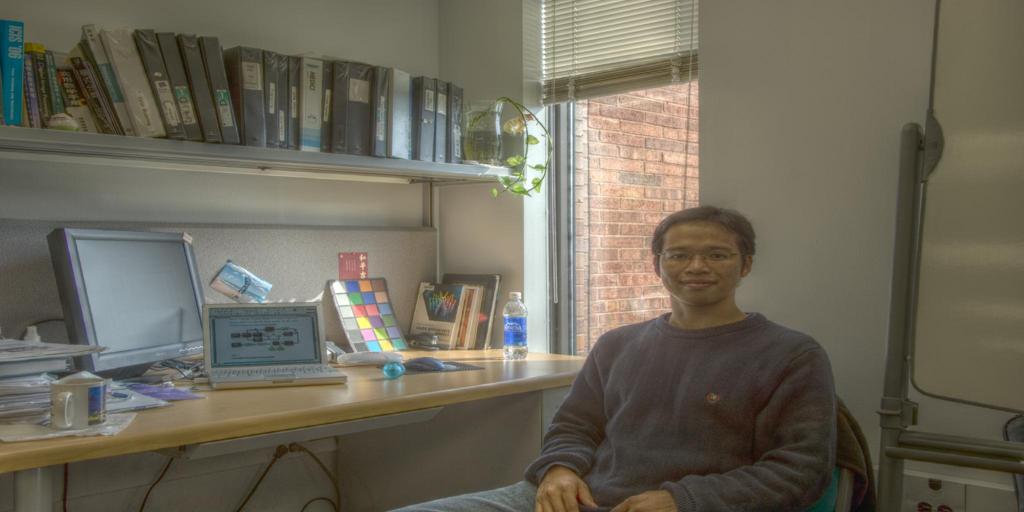}
	\end{subfigure} 	
	\begin{subfigure}[b]{.243\textwidth}
		\caption{ReinhardTMO (0.76)}
		\includegraphics[width=\linewidth]{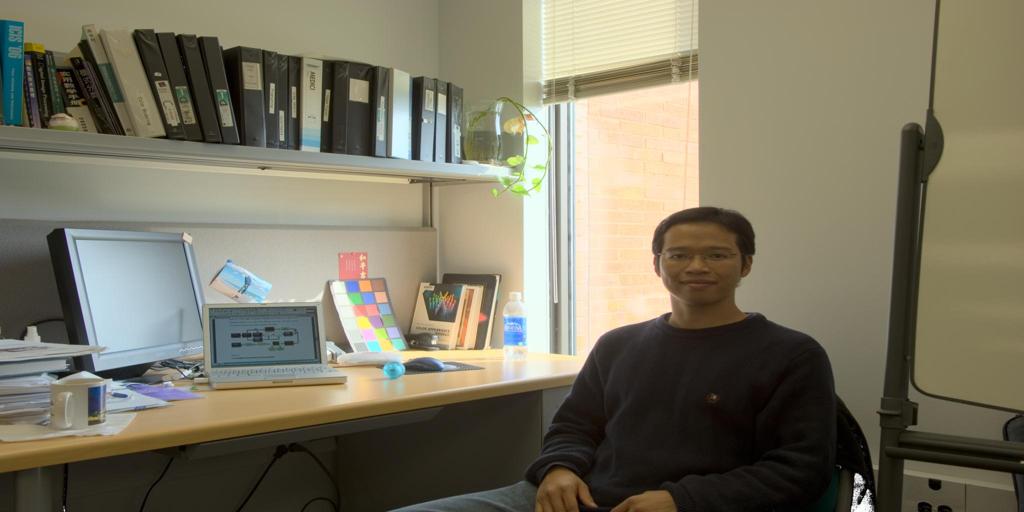}
	\end{subfigure} 

	\begin{subfigure}[b]{.01\textwidth}
		\raisebox{.3in}{\rotatebox[origin=t]{90}{\footnotesize Scene-2}}   
	\end{subfigure}		
	\begin{subfigure}[b]{.243\textwidth}
		\includegraphics[width=\linewidth]{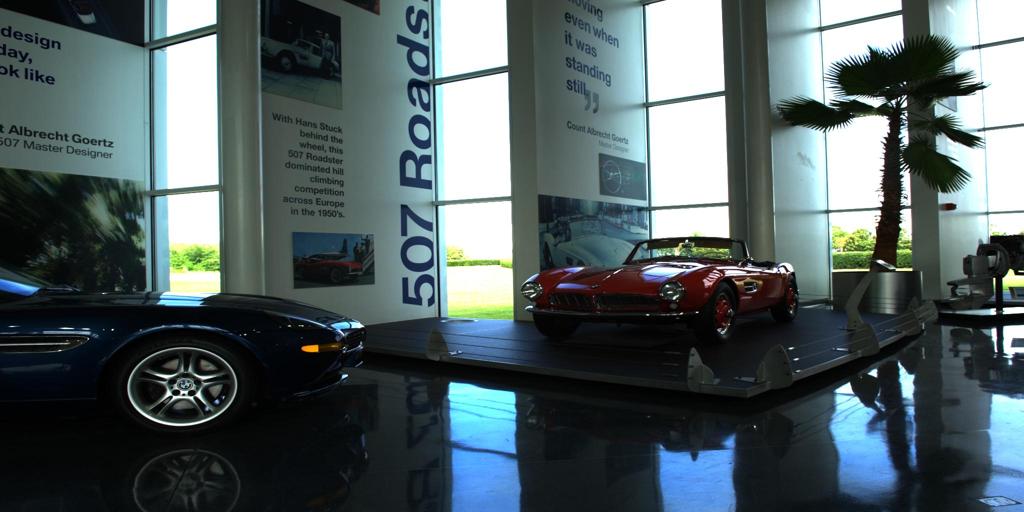}
	\end{subfigure} 
	\begin{subfigure}[b]{.243\textwidth}
		\caption{DeepTMO (0.86)}
		\includegraphics[width=\linewidth]{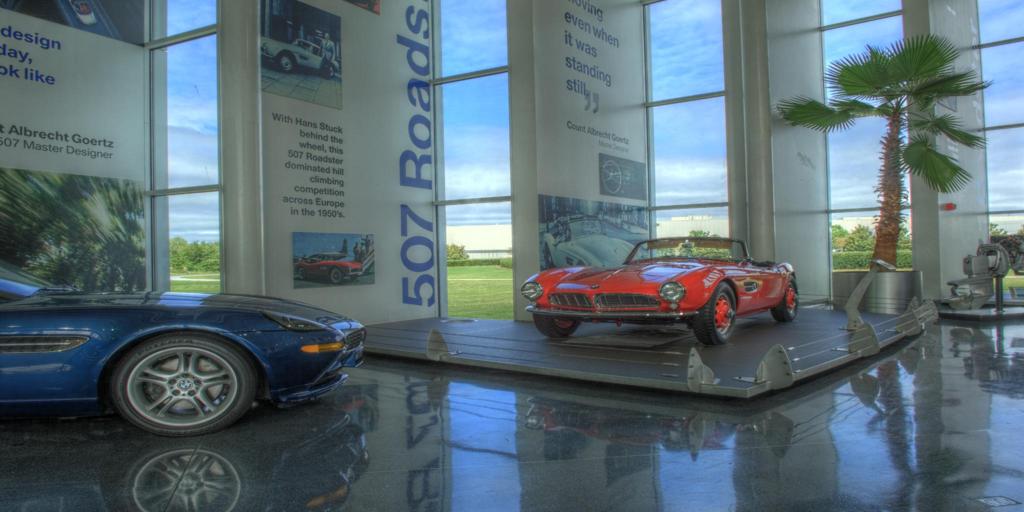}
	\end{subfigure} 	
	\begin{subfigure}[b]{.243\textwidth}
		\caption{ManitukTMO (0.84)}
		\includegraphics[width=\linewidth]{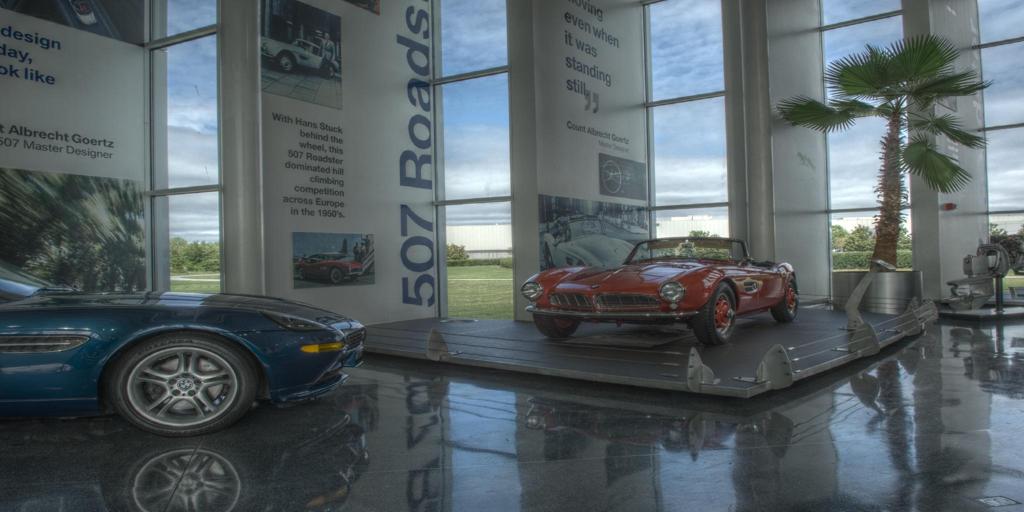}
	\end{subfigure} 	
	\begin{subfigure}[b]{.243\textwidth}
		\caption{DurandTMO (0.80)}
		\includegraphics[width=\linewidth]{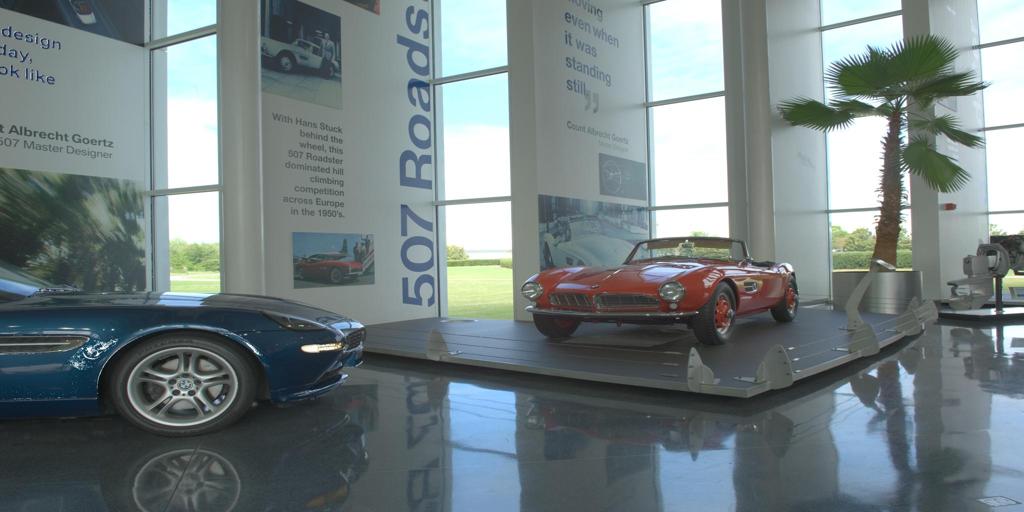}
	\end{subfigure} 
	
	\begin{subfigure}[b]{.01\textwidth}
		\raisebox{.3in}{\rotatebox[origin=t]{90}{\footnotesize Scene-3}}   
	\end{subfigure}				
	\begin{subfigure}[b]{.243\textwidth}
		\includegraphics[width=\linewidth]{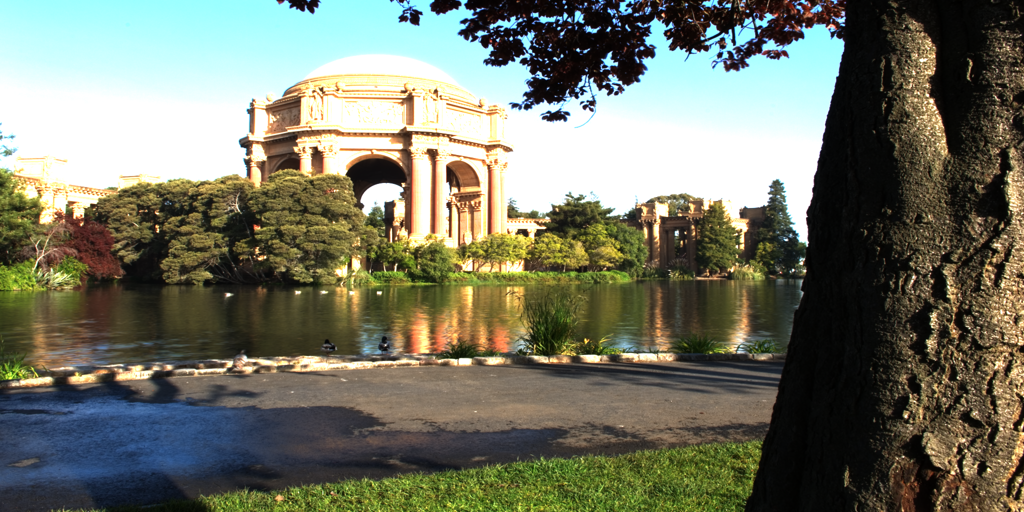}
	\end{subfigure} 
	\begin{subfigure}[b]{.243\textwidth}
		\caption{DeepTMO (0.92)}
		\includegraphics[width=\linewidth]{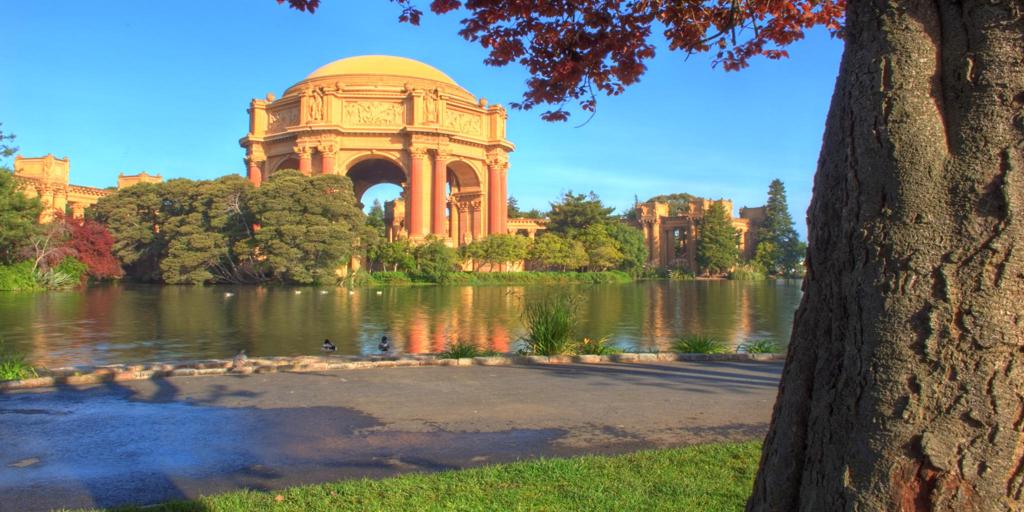}
	\end{subfigure} 	
	\begin{subfigure}[b]{.243\textwidth}
		\caption{ManitukTMO (0.92)}
		\includegraphics[width=\linewidth]{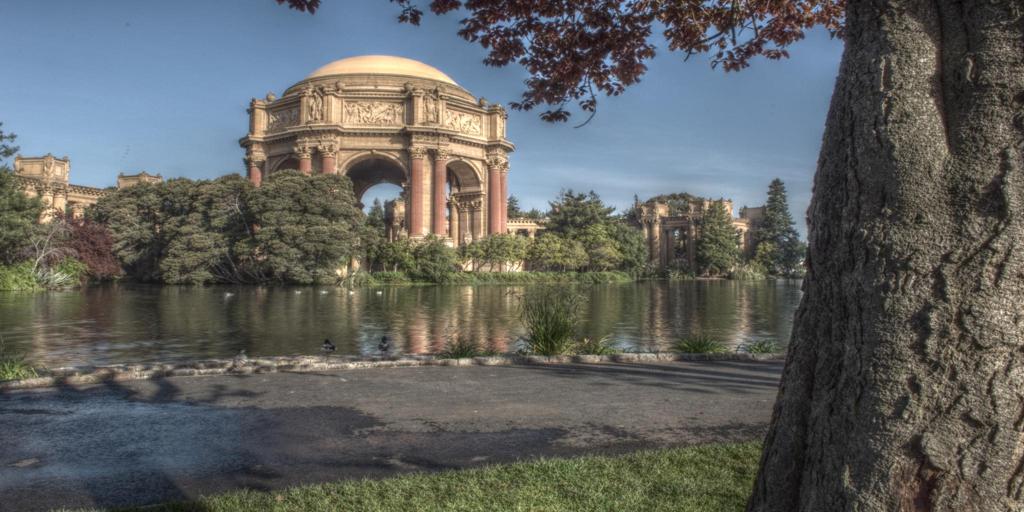}
	\end{subfigure} 	
	\begin{subfigure}[b]{.243\textwidth}
		\caption{DragoTMO (0.87)}
		\includegraphics[width=\linewidth]{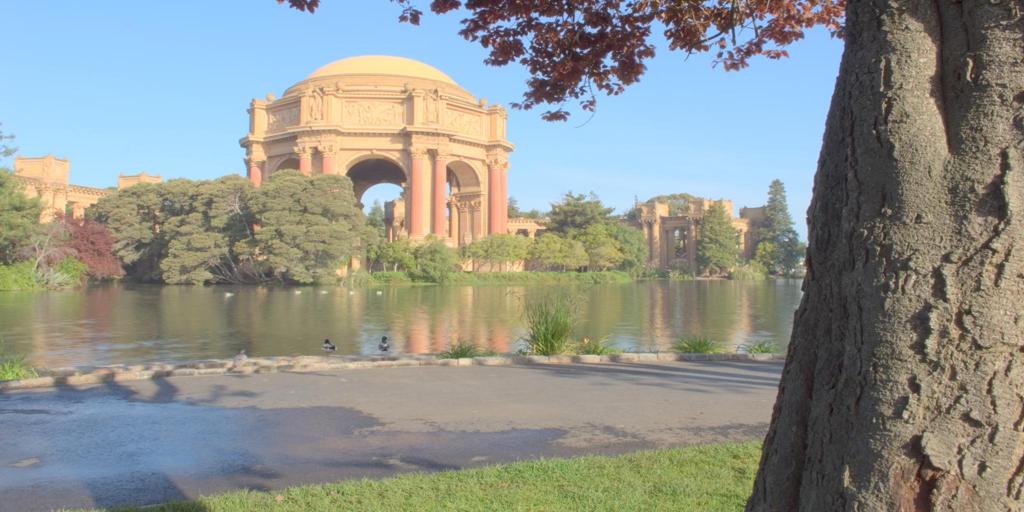}
	\end{subfigure}

	\begin{subfigure}[b]{.01\textwidth}
		\raisebox{.3in}{\rotatebox[origin=t]{90}{\footnotesize Scene-4}}   
	\end{subfigure}			
	\begin{subfigure}[b]{.243\textwidth}
		\includegraphics[width=\linewidth]{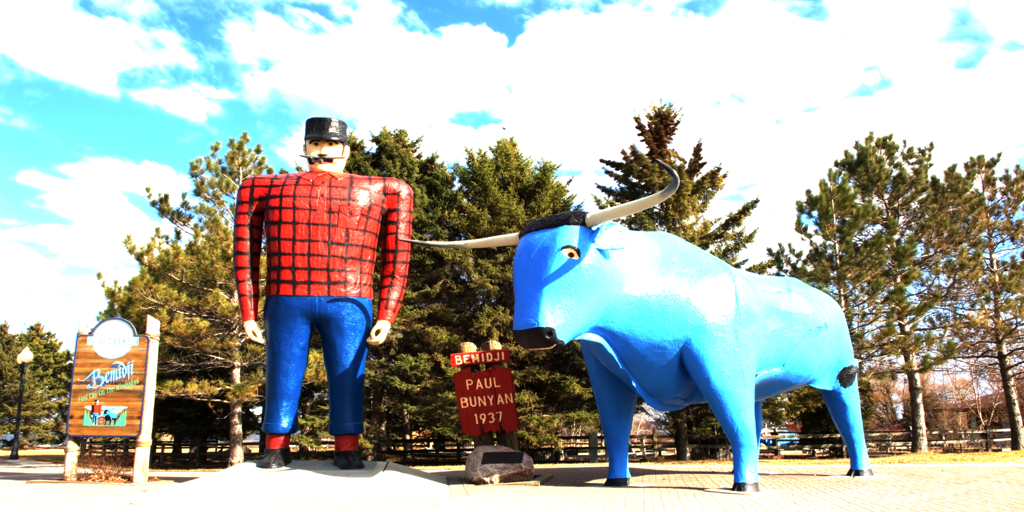}
	\end{subfigure} 
	\begin{subfigure}[b]{.243\textwidth}
	\caption{DeepTMO (0.87)}
		\includegraphics[width=\linewidth]{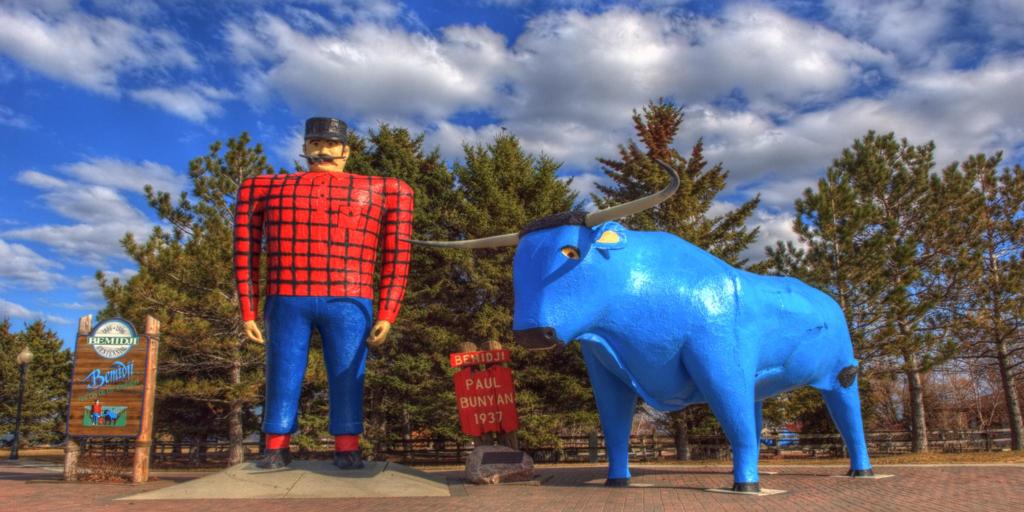}
	\end{subfigure} 		
	\begin{subfigure}[b]{.243\textwidth}
		\caption{FattalTMO (0.88)}
		\includegraphics[width=\linewidth]{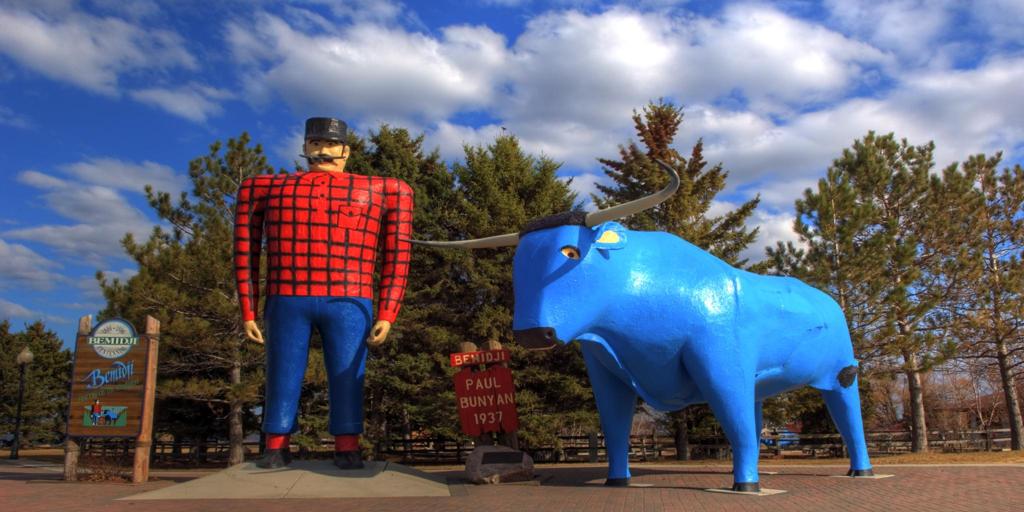}
	\end{subfigure} 	
	\begin{subfigure}[b]{.243\textwidth}
		\caption{SchlickTMO (0.85)}
		\includegraphics[width=\linewidth]{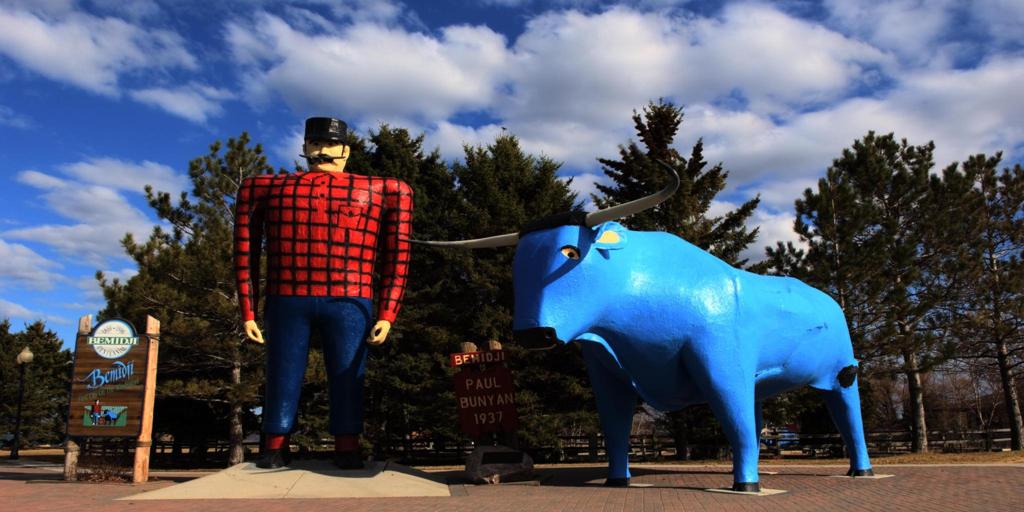}
	\end{subfigure}		
	
	\begin{subfigure}[b]{.01\textwidth}
		\raisebox{.3in}{\rotatebox[origin=t]{90}{\footnotesize Scene-5}}   
	\end{subfigure}			
	\begin{subfigure}[b]{.243\textwidth}
		\includegraphics[width=\linewidth]{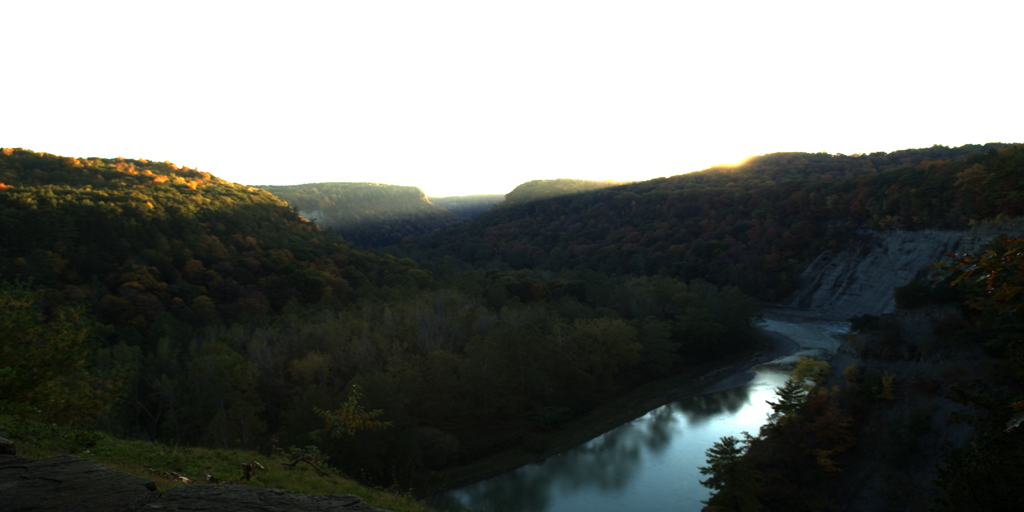}
	\end{subfigure} 
	\begin{subfigure}[b]{.243\textwidth}
	    \caption{DeepTMO (0.76)}
		\includegraphics[width=\linewidth]{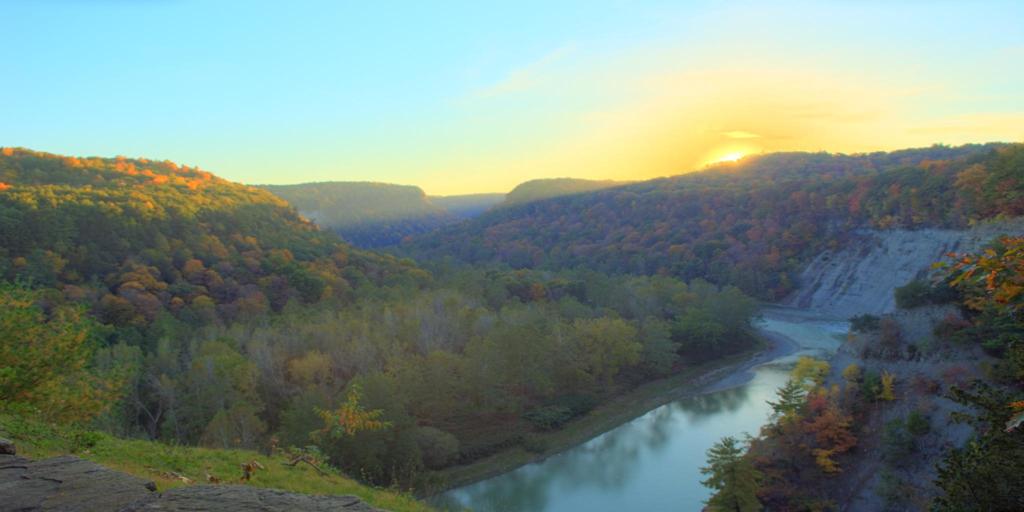}
	\end{subfigure} 		
	\begin{subfigure}[b]{.243\textwidth}
		\caption{FattalTMO (0.73)}
		\includegraphics[width=\linewidth]{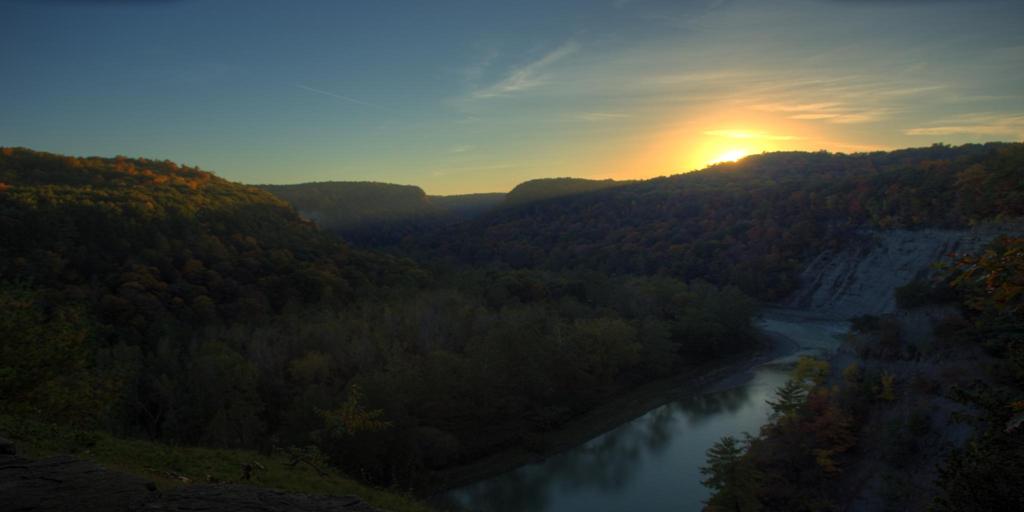}
	\end{subfigure} 	
	\begin{subfigure}[b]{.243\textwidth}
		\caption{SchlickTMO (0.72)}
		\includegraphics[width=\linewidth]{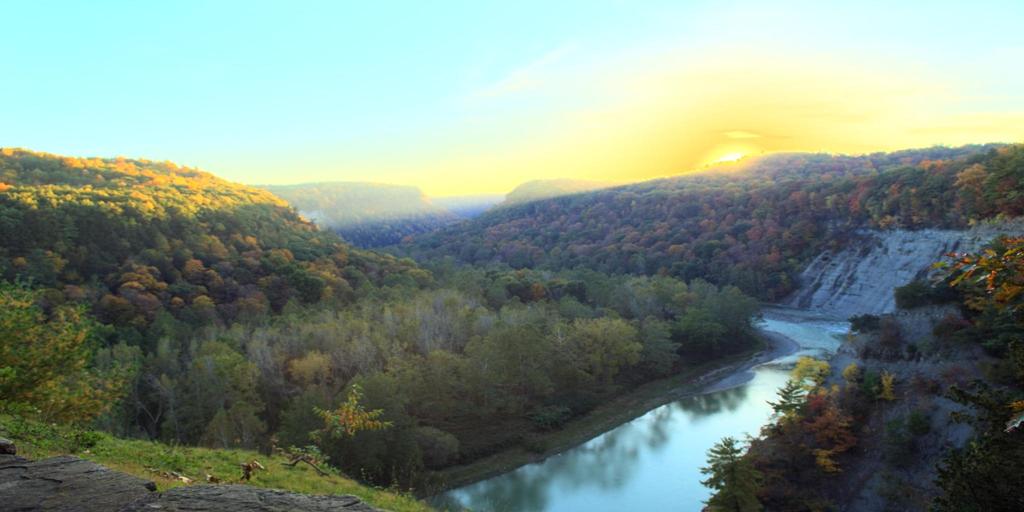}
	\end{subfigure}	
	
	\begin{subfigure}[b]{.01\textwidth}
		\raisebox{.3in}{\rotatebox[origin=t]{90}{\footnotesize Scene-6}}   
	\end{subfigure}			
	\begin{subfigure}[b]{.243\textwidth}
		\includegraphics[width=\linewidth]{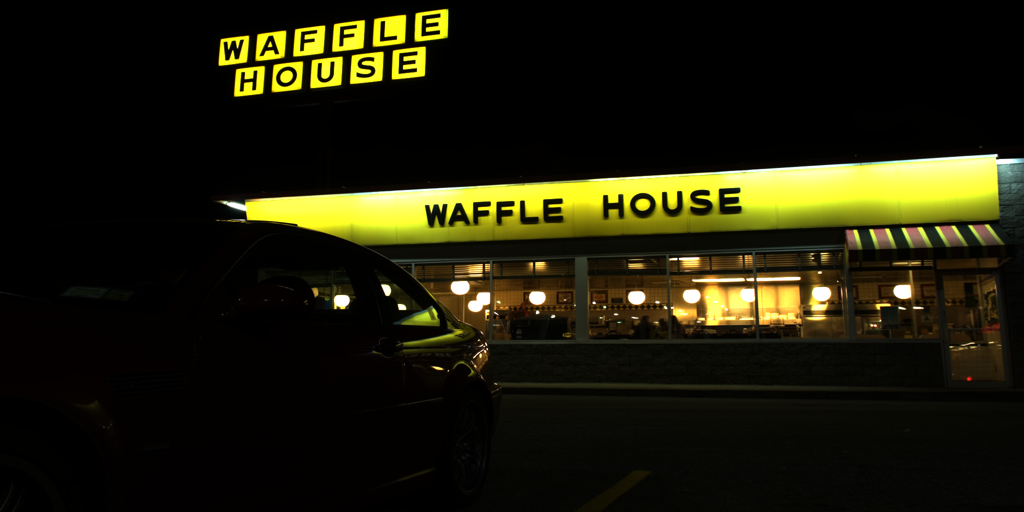}
	\end{subfigure} 
	\begin{subfigure}[b]{.243\textwidth}
	 \caption{DeepTMO (0.88)}
		\includegraphics[width=\linewidth]{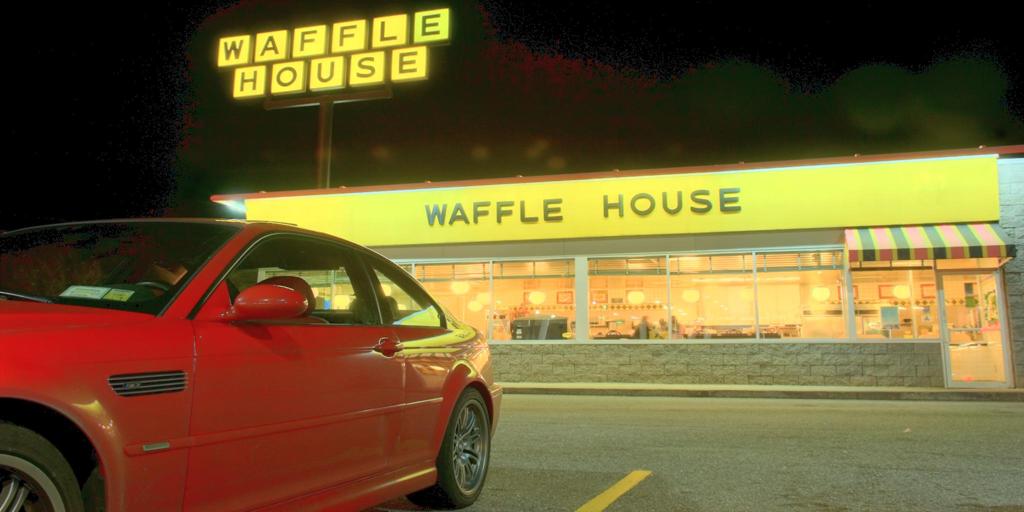}
	\end{subfigure} 		
	\begin{subfigure}[b]{.243\textwidth}
		\caption{PattnaikTMO (0.81)}
		\includegraphics[width=\linewidth]{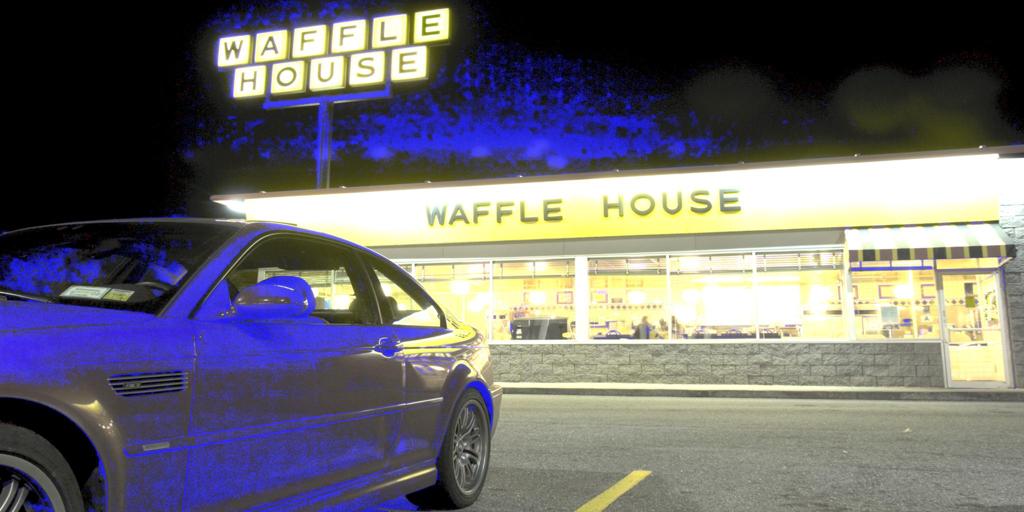}
	\end{subfigure} 	
	\begin{subfigure}[b]{.243\textwidth}
		\caption{MantiukTMO (0.80)}
		\includegraphics[width=\linewidth]{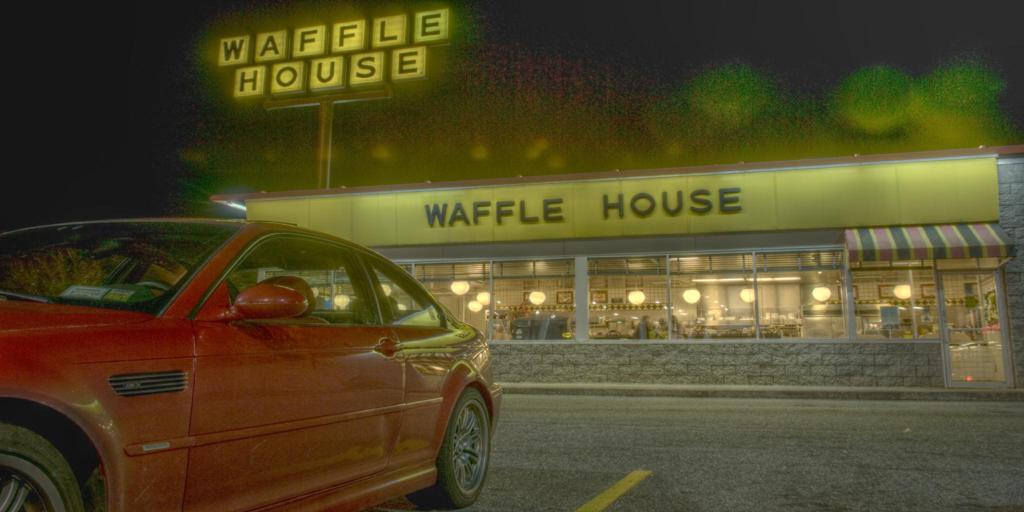}
	\end{subfigure}

	\begin{subfigure}[b]{.01\textwidth}
		\raisebox{.3in}{\rotatebox[origin=t]{90}{\footnotesize Scene-7}}   
	\end{subfigure}		
	\begin{subfigure}[b]{.243\textwidth}
		\includegraphics[width=\linewidth]{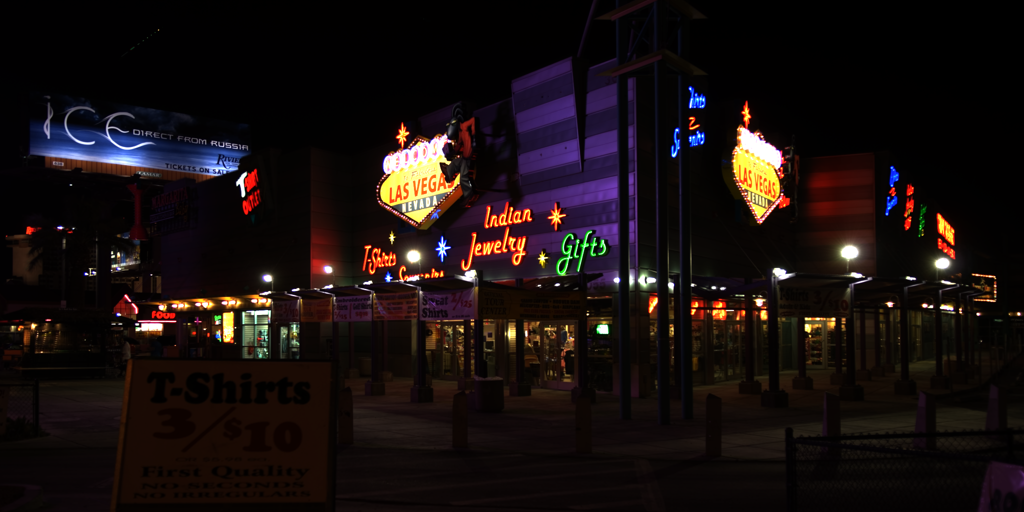}
	\end{subfigure} 
	\begin{subfigure}[b]{.243\textwidth}
	  \caption{DeepTMO (0.87)}
		\includegraphics[width=\linewidth]{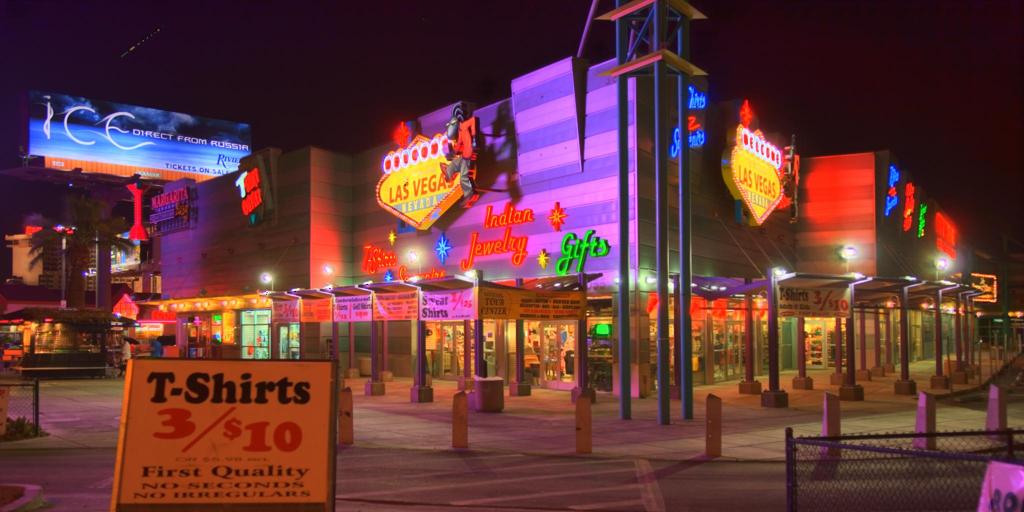}
	\end{subfigure} 		
	\begin{subfigure}[b]{.243\textwidth}
		\caption{ReinhardTMO (0.86)}
		\includegraphics[width=\linewidth]{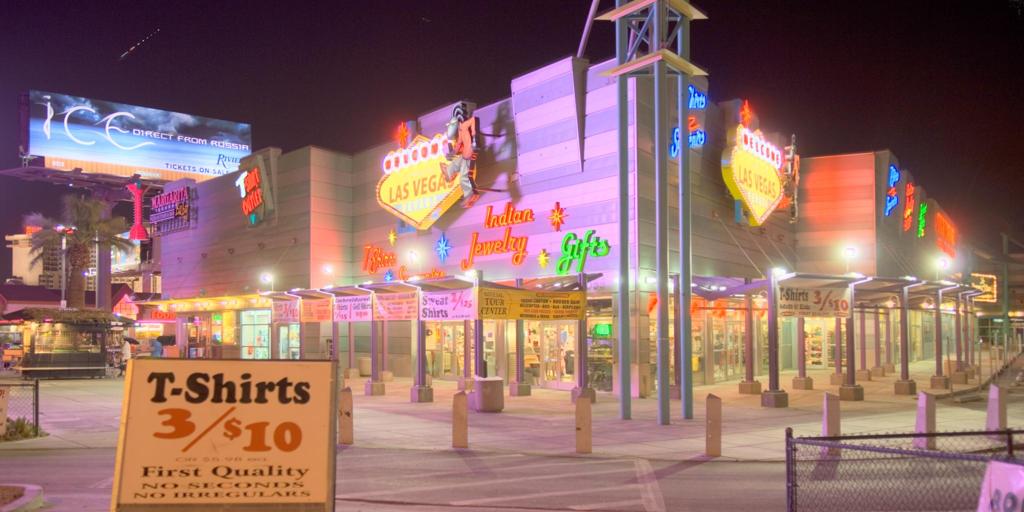}
	\end{subfigure} 	
	\begin{subfigure}[b]{.243\textwidth}
		\caption{DragoTMO (0.83)}
		\includegraphics[width=\linewidth]{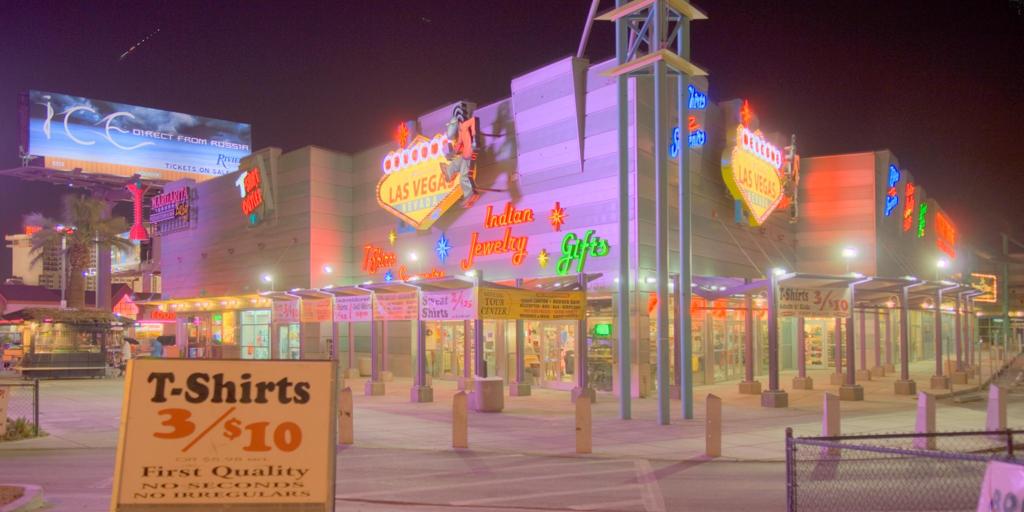}
	\end{subfigure} 			
	\caption{\small Comparison between our DeepTMO outputs and outputs from top-2 ranked tone-mapped scenes on TMQI metrics for a variety of real-world scenes including indoor, scenes with structures, landscape, dark/noisy scenes. In brackets we show corresponding TMQI scores.}\label{fig:QualityScenes}
\end{figure*}

\subsection{Comparison with the Best Quality Tone-Mapped Images}
We begin the comparison of our DeepTMO model against the best quality tone mapped test images to assess the overall capability to reproduce high-quality images over a wide variety of scenes. To obtain the target test image, we follow a similar paradigm as provided in Section~\ref{sec:tmqi}. 

In Fig.~\ref{fig:QualityScenes}, we demonstrate qualitative comparisons of our model with the two top scoring TMOs obtained using TMQI ranking, which includes methods like Mantiuk~\cite{mantiuk06}, Reinhard~\cite{reinhard02}, Fattal~\cite{fattal02}, Durand~\cite{durand02}, Drago~\cite{drago03}, Pattnaik~\cite{patt02} TMO, over 7 explempary real-world scenes representing indoor/outdoor, with humans and structures, in day/night conditions. These sample scenes depict the exemplary mapping of linear HDR content using DeepTMO, where it successfully caters a wide variety of scenes as well as competes with the respective best quality outputs in terms of overall contrast preservation and visual appeal. In scene-1, a scene with human in indoor condition, we observe that our DeepTMO competes closely to the target output while preserving details in under/over exposed areas such as human face, areas under the table or outside the window. Another indoor scene-2, having shiny surfaces (indoor) and saturated outside regions (windows) demonstrate the effectiveness of our model by preserving details in these regions, yielding a high-quality output. Similar observations can be made in outdoor scenes with structures \ie in scene-3 and 4, where we notice that our DeepTMO model effectively tone-maps sharp frequency regions in overly exposed areas such as the dome of the building, the clouds in the sky or the cow's body. Landscape scene-5 has similar observations in the rising sun and dark forest regions. Although multi-scale DeepTMO design pays attention to the global and minute sub-regional information, the preservation of illumination and details in dull and overly bright regions is also due to the presence of the FM-loss term, which in turn utilizes features from different $D$ layers. Since $D$ is focused on localized image-patches, the FM-term implicitly understands how to compress or enhance luminance in specific regions. 
 
More interestingly, we observe that DeepTMO suppresses noisy disturbances (\ie above the Waffle House store) in dark scene-6, which appears more pronounced in the two best performing tone-mapped images. This can be reasoned owing to the addition of VGG and FM-loss terms which guides the network to handle the noisy repetitive patterns and dark sensor-noise while preserving the natural scene statistics. Furthermore, we showcase a night time high-contrast scene-7, where our DeepTMO competes closely with the two best quality outputs while preserving the overall contrast ratio. 
However, we do observe the images obtained with our method have more saturated colors which we discuss later in Section~\ref{sec:limit}.
 
\begin{figure}
	\centering
	\captionsetup[subfigure]{labelformat=empty}		
	\begin{subfigure}[b]{.01\textwidth}
		\raisebox{.7in}{\rotatebox[origin=t]{90}{\footnotesize{\cite{durand02} (.90) ----- DeepTMO (0.92)}}}   
	\end{subfigure}		
	\begin{subfigure}[t]{0.43\textwidth}
		\includegraphics[width=\textwidth,height=.5\textwidth]{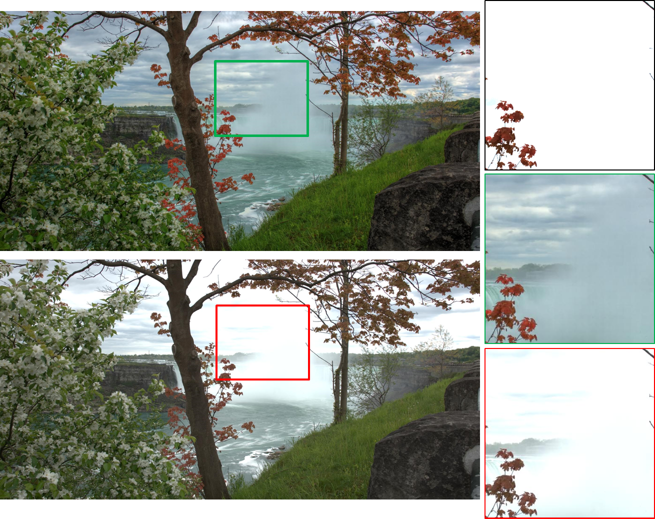}
	\end{subfigure}	
	\begin{subfigure}[b]{.01\textwidth}
		\raisebox{1.2in}{\rotatebox[origin=t]{270}{\footnotesize{HDR-Linear}}}   
	\end{subfigure}		
	\vspace{1 mm}
	
	\begin{subfigure}[b]{.01\textwidth}
		\raisebox{.7in}{\rotatebox[origin=t]{90}{\footnotesize{\cite{mantiuk06} (0.7) ----- DeepTMO (0.82)}}}   
	\end{subfigure}		
	\begin{subfigure}[t]{0.43\textwidth}
		\includegraphics[width=\textwidth,height=.5\textwidth]{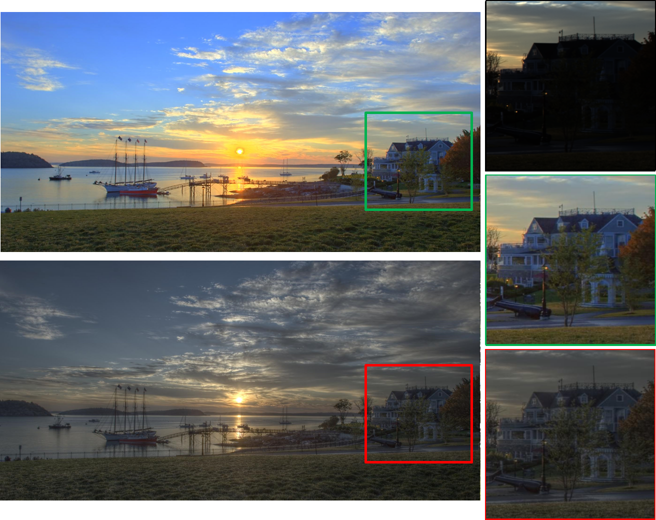}
	\end{subfigure}	
	\begin{subfigure}[b]{.01\textwidth}
		\raisebox{1.2in}{\rotatebox[origin=t]{270}{\footnotesize{HDR-Linear}}}   
	\end{subfigure}		
	
	
	\caption{\small Comparison between DeepTMO and targets, highlighting the zoom-ins with the corresponding HDR-linear input.
	}\label{fig:withbestTMO}
\end{figure}

Though in most cases our DeepTMO competes well with target images, in some cases we observe that it even outperforms them with respect to TMQI scores. Fig.~\ref{fig:withbestTMO} compares two exemplary HDR scenes from the test dataset that are mapped using the DeepTMO and their corresponding target TMOs in day and evening time-settings. In the first row, DeepTMO successfully preserves the fine details in the sky along with the waterfall and the mountains in the background. For a darker evening scene in second row, DeepTMO compensates the lighting and preserves the overall contrast of the generated scene. Even though we observe a halo ring around sun using our method (which we analyze later in Section~\ref{sec:limit}), our TMQI score is considerably higher mainly because the TMQI metric is color-blind. 

One possible explanation of such outcomes is the ability of the generator to learn the manifold of all available best tone mapping operators and subsequently developing a superior tone mapping functionality (from this manifold), which yields optimal output depending upon the scene. In other words, this manifold learning can be observed as a loose formulation built over the ideal characteristics (both global and local) desired for tone-mapping of different scene-types present in the training-set. In fact, learning such a complex mapping functionality is non-trivial by using a global TMQI metric score alone. This further confirms the goal of our training strategy.

\begin{figure}[h]
	\centering
	\begin{subfigure}[t]{0.475\textwidth}
		\includegraphics[width=\textwidth, height= 0.6\textwidth]{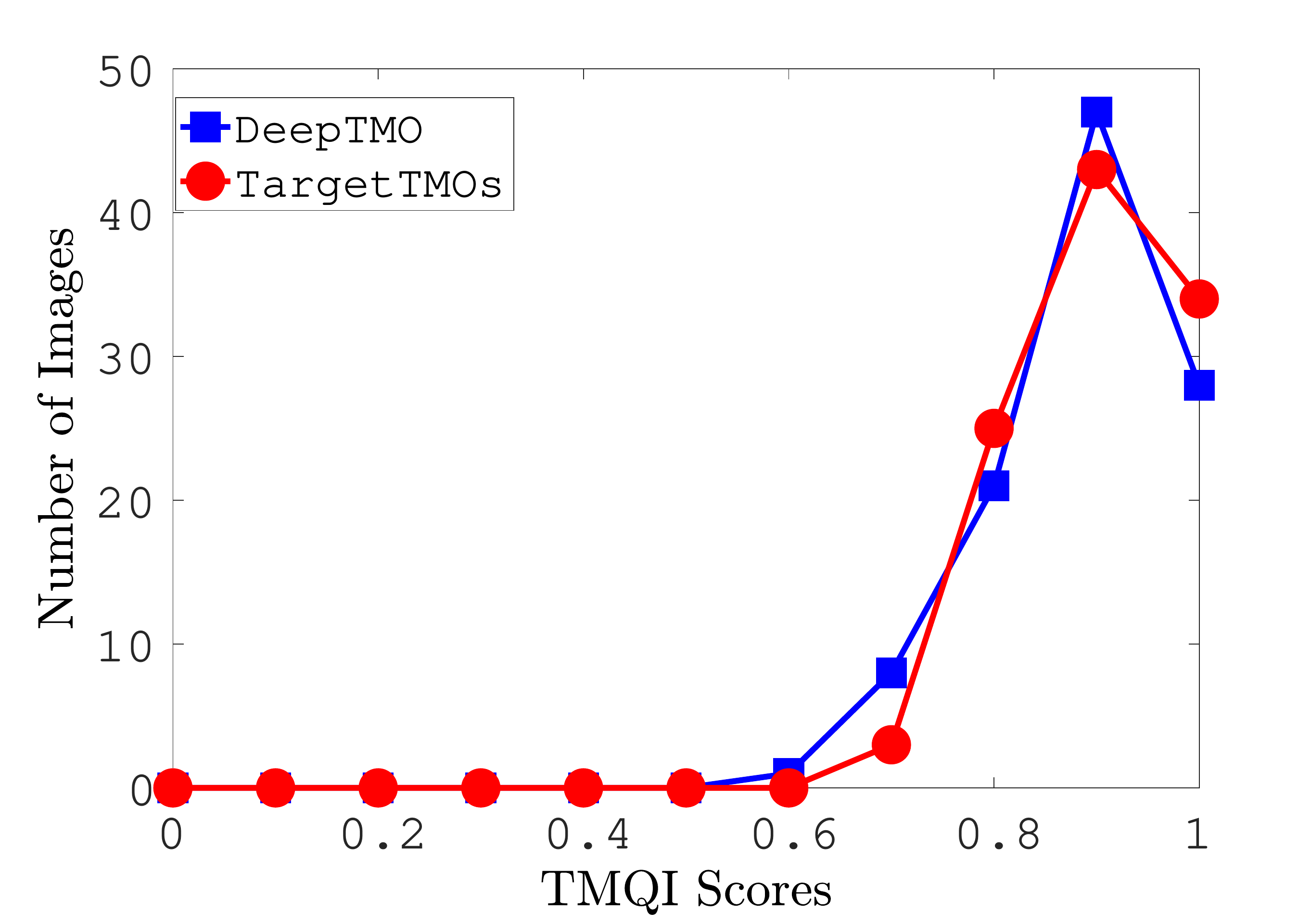}
	\end{subfigure}	
	\caption{\small Quantitative performance comparison of best performing DeepTMO with the target TMOs.}\label{fig:CurveswithbestTMO}
\end{figure}

\paragraph{Quantitative Analysis}
To further demonstrate the high-quality mapping capability of DeepTMO models on all the 105 real world scenes, in Fig.~\ref{fig:CurveswithbestTMO}, we show a distribution plot of the number of scenes against the TMQI Scores. For completeness, we also provide scores achieved by target tone-mapped outputs. The curves clearly show that the generated tone mapped images for DeepTMO compete closely with the best available tone mapped images on the objective metrics with DeepTMO fairing the best amongst all.

We provide quantitative analysis in Table~\ref{tab:tmqi}, to showcase the performance of our proposed model with the existing approaches. For each method, the TMQI scores are averaged over 105 scenes of the test dataset. The final results show that our proposed tone mapping model adapts for the variety of scenes and hence, achieves highest score. Please note that standard TMOs were applied with default parameter settings and hence results may improve for them by parameter optimization. Still performance of our fully automatic approach is highly competitive.

\begin{table}
	\caption{\small \textit{Quantitative Results}. Mean TMQI scores on the test-set of 105 images.}\label{tab:tmqi}
	\centering
	\begin{tabular}{cc}
		\centering
		TMOs  & TMQI  \\ 
		\hline
		Tumblin~\cite{tumblin99} TMO & 0.69 $\pm 0.06$ \\
	    Chiu~\cite{chiu96} TMO & 0.70 $\pm 0.05$ \\
	    Ashikh~\cite{ashikhmin02} TMO  & 0.70 $\pm 0.06$ \\
		Ward~\cite{ward97} TMO & 0.71 $\pm 0.07$  \\
		Log~\cite{Banterle} TMO & 0.72 $\pm 0.09$ \\
		Gamma~\cite{Banterle} TMO  & 0.76 $\pm 0.07$ \\
		Pattnaik~\cite{patt02} TMO  & 0.78 $\pm 0.04$ \\
		Schlick~\cite{schlick94} TMO & 0.79 $\pm 0.09$ \\
		Durand~\cite{durand02} TMO & 0.81 $\pm 0.10$\\
		Fattal~\cite{fattal02} TMO & 0.81 $\pm 0.07$ \\
		Drago~\cite{drago03} TMO & 0.81 $\pm 0.06$ \\
		Reinhard~\cite{reinhard02} TMO & 0.84 $\pm 0.07$ \\
		Mantiuk~\cite{mantiuk06} TMO & 0.84 $\pm 0.06$ \\
		DeepTMO (Single G - Single G)& 0.79 $\pm 0.06$ \\
		DeepTMO (Single G - Multi D)& 0.81 $\pm 0.05$ \\
		DeepTMO (Multi G - Single D)& 0.80 $\pm 0.07$ \\
		DeepTMO (Multi G - Multi D)& \textbf{0.88} $\pm \textbf{0.06}$ \\
	\end{tabular}
\end{table}

\begin{figure}
	\centering		
	\begin{subfigure}[t]{0.38\textwidth}
		\includegraphics[width=\textwidth, height= 0.6\textwidth]{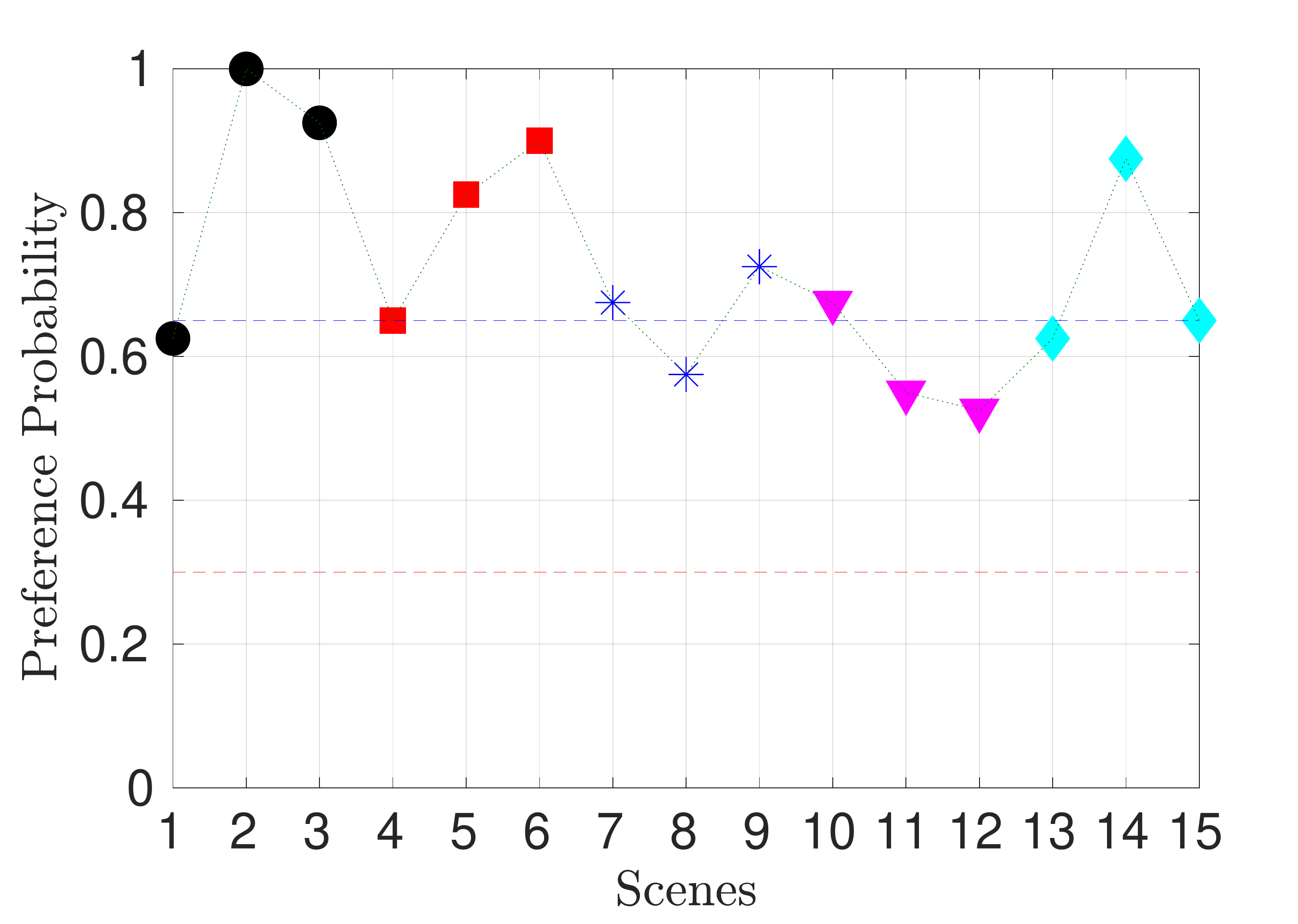}
	\end{subfigure}					
	\begin{subfigure}[t]{0.08\textwidth}
		\includegraphics[width=\textwidth]{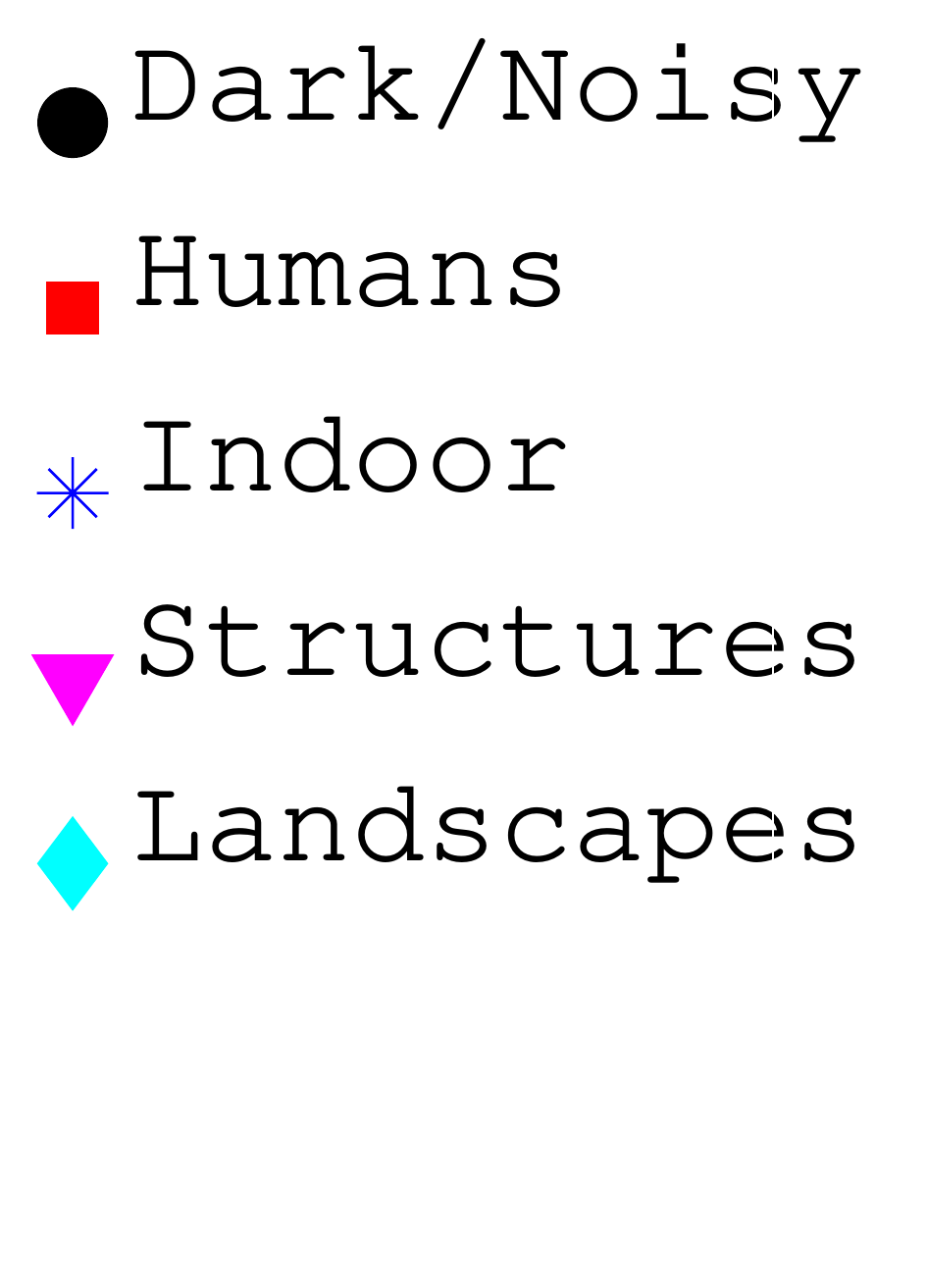}
	\end{subfigure}							
	\caption{\small Subjective Test Results. Preference probability of our DeepTMO over best performing target TMOs for 15 scenes representing 5 different scene categories.}\label{fig:subjectiveTests}			
\end{figure}
\begin{figure}[t]
	\centering
	\begin{adjustbox}{width=.48\textwidth}
		\begin{tikzpicture}
		\begin{axis}[
		xbar,
		bar width=0.09cm,
		axis x line       = none,
		y axis line style={opacity=0},
		width=5cm, height=2.2cm,
		tickwidth = 1pt,
		ytick=data,
		yticklabel style = {font=\tiny},
		symbolic y coords = {Single-G-Single-D,Multi-D-Single-G,Multi-G-Single-D,Multi-G-Multi-D},
		nodes near coords, nodes near coords align={horizontal},{font=\tiny},
		]
		\addplot[red!20!black,fill=red!80!white]  coordinates { (0.01879,Single-G-Single-D) (0.02,Multi-D-Single-G) (0.0201,Multi-G-Single-D) (0.02097,Multi-G-Multi-D)};
		\end{axis}
		\end{tikzpicture}
	\end{adjustbox}
	\caption{\textit{Computation time in seconds.}}\label{fig:computation}
\end{figure}
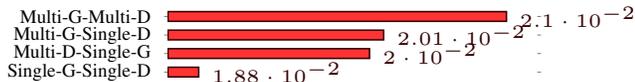

\paragraph{Computation Time}
Inference is performed on test-images of size $1024 \times 2048$ and takes on an average 0.0187 sec. for single-scale and 0.0209 sec. for multi-scale designs, as shown in Figure~\ref{fig:computation}. 

\subsection{Quality Evaluation}
We performed a subjective pairwise comparison to validate the perceived quality of our tone-mapped images. 20 people participated in this subjective study, with age range of 23-38 years, normal or corrected-to-normal vision. 
\subsubsection{Test Environment and Setup}
The tests were carried out in a room reserved for professional subjective tests with ambient lighting conditions. A Dell UltraSharp 24 Monitor (DELL U2415) was used for displaying images with screen resolution $1920\times1200$ at $59$ hz. The desktop background window was set at $128$ gray value.

Each stimuli included a pair of tone mapped images for a given scene, where each pair always consisted of an image produced by DeepTMO and the other one obtained using the best-performing tone mapping functions based on the TMQI rankings. To cater a wide variety of content, we selected 15 scenes from 105 test-set images, representing 5 different categories (3 scenes per category) namely, i) Humans, ii) Dark/Noisy, iii) Indoor, iv) Structures, and v) Landscapes.  

\subsubsection{Procedures}  
We conducted a pair-wise subjective experiment where the observer was asked to choose an image by showing a pair of images side-by-side. The option same was not included to force users to choose one of the stimuli. Each participant was asked to select an image which is more realistic and appealing to him/her. Participants were provided with unlimited time to make their decision and record their choice. The experiment was divided into a training and test session, where training involved each participant being briefed to familiarize with the subjective quality evaluation task. Each observer compared a pair of image twice, having each tone-mapped image displayed on both sides (\eg DeepTMO  vs. first-best tone mapped and  first-best tone mapped vs. DeepTMO).

\subsubsection{Results}
In order to quantify the results of pairwise subjective tests, we scaled the winning frequencies of the model to the continuous quality-scores using the widely known Bradley-Terry (BT) model in~\cite{BradleyTerry}. The scaling is performed using the statistical analysis proposed in~\cite{Hanhart} to determine whether the perceived visual quality difference of the compared models is statistically significant. The preference probability for our method $Pref-Prob_{(DeepTMO)} $ is mathematically given as: 
\begin{equation}
Pref-Prob_{(DeepTMO)} = \frac{w_{DeepTMO}}{N} + \frac{t}{2\cdot N}
\end{equation}

where $w_{DeepTMO}$ is the winning frequency of our proposed model, $t$ is the tie frequency and $N$ is the total number of participants. The statistical model relies on the hypothesis that each compared TMO in the pairwise test shares equal probability of occurrence \ie 0.5 and hence, follows a Binomial distribution. Based on the initial hypothesis, a Binomial test was performed on the collected data and the critical thresholds were obtained by plotting the cumulative distribution function of the Binomial distribution. By setting $95\%$ as the level of significance, if we receive 13 $(B(13,20,0.5) = 0.9423)$ or more votes for our proposed method, we consider our tone-mapped image to be significantly favored in terms of subjective quality. Similarly, by setting $5\%$ as the significance level, if we receive 6  $(B(6,20,0.5) = 0.0577)$ or less votes for our proposed method, we consider our tone-mapped image to be least favored in terms of subjective quality.

The results of the pair-wise subjective quality experiment are shown in Fig.~\ref{fig:subjectiveTests}. The two lines (blue and red) mark probabilities of high ($13/20 = .65$) and low ($6/20 =.30$) favor-abilities respectively. Looking at the results, we observe that DeepTMO images have been significantly preferred over best TMQI rated tone mapped images for most of the scenes, for different possible categories. In general, we observed that subjects preferred our tone-mapped LDR scenes which preserve the contrast well. Based on some informal post-experiment interviews, we found that best TMQI rated target images, preserving fine details were least realistic and more like paintings to observers. A small set of images used in subjective tests is shown in Fig.~\ref{fig:QualityScenes}.

	\section{Conclusion, Limitations and Future work}
 \begin{figure}
 	\centering	
 	\captionsetup[subfigure]{}				
 	
 	\begin{subfigure}[t]{0.115\textwidth}
 		\includegraphics[width=\textwidth,height=.9\textwidth]{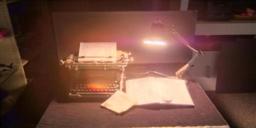}
 		\caption{}
 	\end{subfigure}					
 	\begin{subfigure}[t]{0.115\textwidth}
 		\includegraphics[width=\textwidth,height=.9\textwidth]{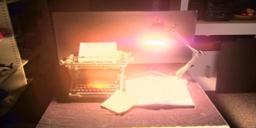}
 		\caption{}
 	\end{subfigure}		
 	\begin{subfigure}[t]{0.115\textwidth}
 		\includegraphics[width=\textwidth,height=.9\textwidth]{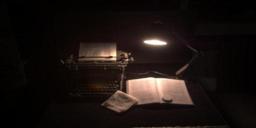}
 		\caption{}
 	\end{subfigure} 
 	\begin{subfigure}[t]{0.115\textwidth}
 		\includegraphics[width=\textwidth,height=.9\textwidth]{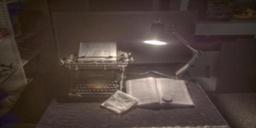}
 		\caption{}
 	\end{subfigure} 	
 	\caption{\small Top TMQI scoring TMOs showing not-so-visually desirable outputs. (a) DeepTMO output, (b), (c) and (d) are 3 top ranking TMO output. }\label{fig:tmqi-case}			
 \end{figure} 
 
 \begin{figure}
 	\centering	
 	\captionsetup[subfigure]{}				
 	
 	\begin{subfigure}[t]{0.115\textwidth}
 		\includegraphics[width=\textwidth,height=.9\textwidth]{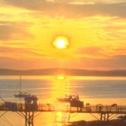}
 		\caption{}
 	\end{subfigure}					
 	\begin{subfigure}[t]{0.115\textwidth}
 		\includegraphics[width=\textwidth,height=.9\textwidth]{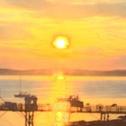}
 		\caption{}
 	\end{subfigure}		
 	\begin{subfigure}[t]{0.115\textwidth}
 		\includegraphics[width=\textwidth,height=.9\textwidth]{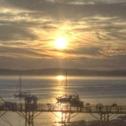}
 		\caption{}
 	\end{subfigure} 
 	\begin{subfigure}[t]{0.115\textwidth}
 		\includegraphics[width=\textwidth,height=.9\textwidth]{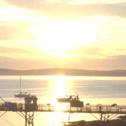}
 		\caption{}
 	\end{subfigure} 
 	
 	\caption{\small Halo effect. (a) DeepTMO output, (b) DeepTMO trained with log-scaled values, (c) and (d) 2 top ranking TMO outputs. }\label{fig:haloeffect}			
 \end{figure}  
Designing a fast, automated tone-mapping operator that can reproduce best subjective quality outputs from a wide range of linear-valued HDR scenes is a daunting task. Existing TMOs address some specific characteristics, such as overall contrast ratio, local fine-details or perceptual brightness of the scene. However, the entire process of yielding high-quality tone-mapped output remains a time-consuming and expensive task, as it requires an extensive parameter tuning to produce a desirable output for a given scene.

To this end, we present an end-to-end parameter-free DeepTMO. Tailored in a cGAN framework, our model is trained to output realistically looking tone-mapped images, that duly encompass all the various distinctive properties of the available TMOs. We provide an extensive comparison among various architectural design choices, loss functions and normalization methods, thus highlighting the role that each component plays in the final reproduced outputs. Our DeepTMO successfully overcomes the frequently addressed blurry or tiling effects in recent HDR related works~\cite{Gabriel,deepreverse}, a problem of significant interest for several high-resolution learning-based graphical rendering applications as highlighted in ~\cite{Gabriel}. 
By simply learning an HDR-to-LDR cost function under a multi-scale GANs framework, DeepTMO successfully preserves desired output characteristics such as underlying contrast, lighting and minute details present in the input HDR at the finest scale. Lastly, we validate the versatility of our methodology through detailed quantitative and qualitative comparisons with existing TMOs.
  \begin{figure}[h]
  	\centering	
  	\captionsetup[subfigure]{}				
  	\begin{subfigure}[t]{0.145\textwidth}
  		\includegraphics[width=\textwidth]{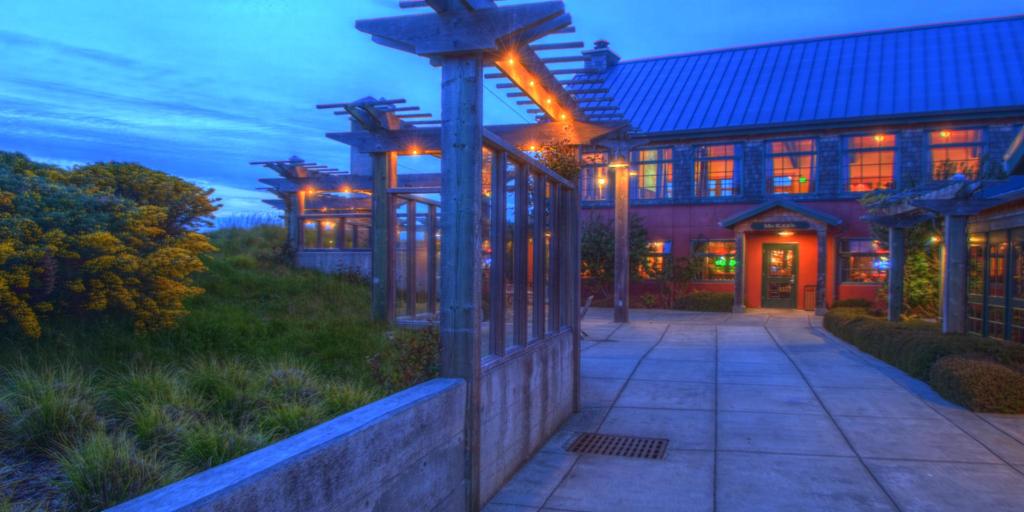}
  		\caption{DeepTMO}
  	\end{subfigure}					
  	\begin{subfigure}[t]{0.145\textwidth}
  		\includegraphics[width=\textwidth]{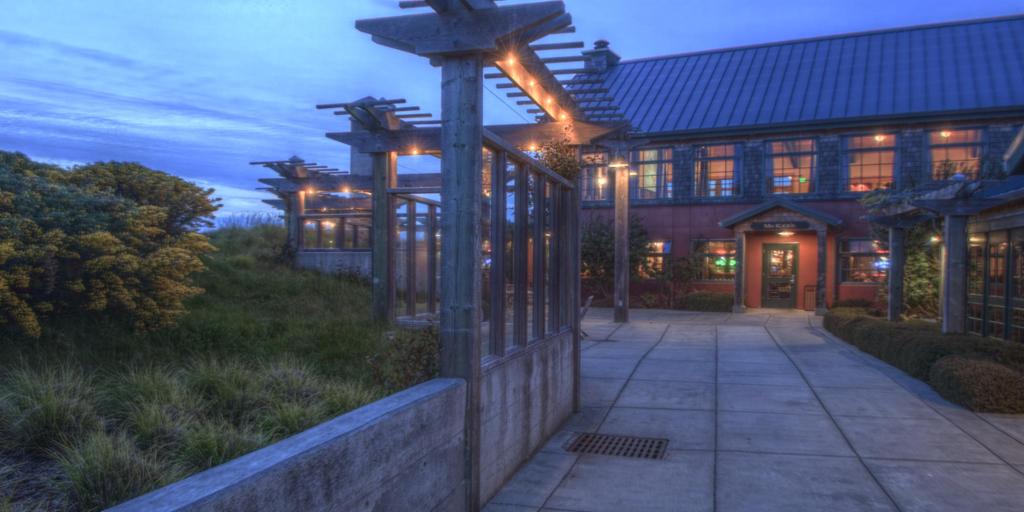}
  		\caption{s=0.5}
  	\end{subfigure}		
  	\begin{subfigure}[t]{0.145\textwidth}
  		\includegraphics[width=\textwidth]{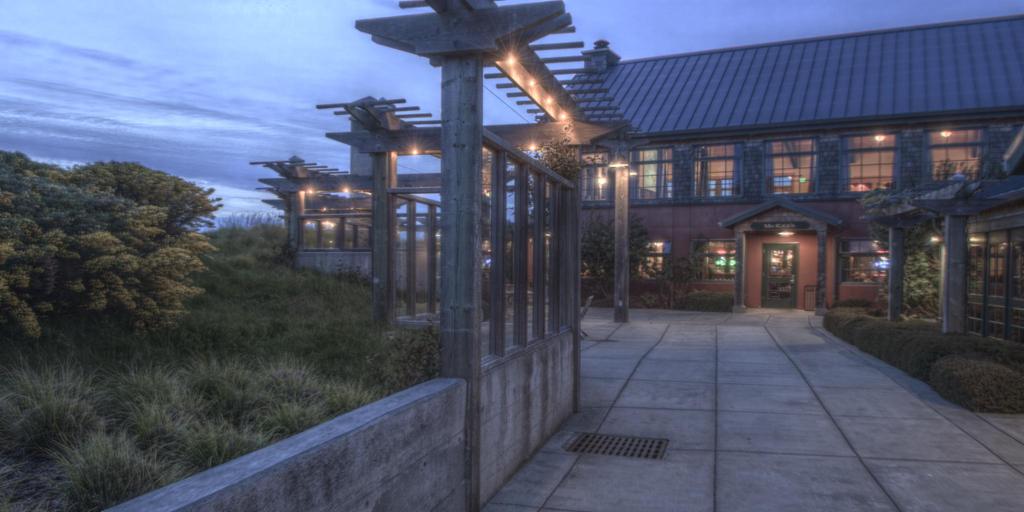}
  		\caption{s=0.3}
  	\end{subfigure}  
  	\caption{\small Color Correction. (a) DeepTMO, (b) and (c) are the color corrected DeepTMO controlled by parameter $s$ from~\cite{colorcorrection}. }\label{fig:CC}			
  \end{figure}
\subsection{Limitations and Future Work}\label{sec:limit} 
\paragraph{Target Selection}
Though DeepTMO successfully demonstrates versatility in addressing wide variety of scenes, its expressive power is limited by the amount of available training data and quality of its corresponding `target'. As noted in Section~\ref{sec:intro}, due to unavailability of subjectively annotated `best tone mapped images' for HDR scenes, we resort to an objective TMQI metric to build the corresponding target LDR. However, the metric itself is not as perfect as the human visual system. We illustrate this point in Fig.~\ref{fig:tmqi-case}. The images ranked lower by TMQI metric in column 3 and 4 are somehow more interesting than their best-ranked counterpart in column 2. 
Such samples can eventually restrict the generation power of our model.
 
Another specific case includes `Halo' artifacts or rings around high illumination regions such as the sun shown in Fig.~\ref{fig:haloeffect}, where DeepTMO (column 1) is compared with the top TMQI scoring outputs in column 3 and column 4. This is mainly due to the inadequate amount of training data consisting of such samples, and the presence of their overly saturated `target' counterparts. As a result, $D$ has very little information about effectively tone-mapping such regions, and thus is unable to guide $G$ to effectively eradicate such effects at generation time. 
To handle such artifacts, we additionally experimented using a log-scale input (column 2) where we observe that even log-scale values do no rectify such effects, thus necessitating the need of adequate training samples.


An alternative future work to address this problem, can be to weakly rely on these `noisy' tone-mapped ground truths images by utilizing a weakly supervised learning paradigm \cite{weaklysupervised}. We can also learn HDR-to-LDR mapping in a completely unsupervised fashion without giving any input-output pairs  \cite{cyclegan}. This would allow the network to decide by itself which is the best possible tone-mapped output simply by independently modeling the underlying distribution of input HDR and output tone mapped images.
%

\paragraph{Color Correction}
Color is an important aspect while rendering high quality subjective tone-mapped outputs. Our proposed method has been trained for efficient luminance compression in HDR scenes and uses the classical color ratios to produce the resulting tone-mapped outputs. Although it provides best subjective quality outputs in most cases, it sometimes can result into overly saturated colors which might look unnatural and perceptually unpleasant. One simple solution could be to simply plug-in existing color correction methods~\cite{colorcorrection} to obtain the desired output. An example is shown in Fig.~\ref{fig:CC}, where color correction has been carried out using the method as proposed in ~\cite{colorcorrection}, which is given by $C_{out} = ((\frac{C_{in}}{L_{in}} - 1)\cdot s +1) \cdot L_{out} $, where s is the color saturation control. Alternately, another interesting solution could be to learn a model to directly map the content from HDR color-space to an LDR colored tone mapped output.

\bibliographystyle{IEEEtran}
\bibliography{sample-bibliography}	
	
	%
	
%
%
%
%
\end{document}